# Spread-F occurrence during geomagnetic storms near the southern crest of the EIA in Argentina

Gilda de Lourdes González[1,2]

[1]Universidad Nacional de Tucumán (UNT)
Av. Independencia 1800 – San Miguel de Tucumán (4000) – Argentina
[2]Universidad del Norte "Santo Tomás de Aquino" (UNSTA)
9 de Julio 165, T4000 San Miguel de Tucumán, Tucumán
gilda.gonzalez@unsta.edu.ar

---

**Abstract** This work presents, for the first time, the analysis of the occurrence of ionospheric irregularities during geomagnetic storms at Tucumán – Argentina, a low latitude station in the Southern American longitudinal sector ( 26.9 ° S, 294.6 ° E; magnetic latitude 15.5 ° S), near the southern crest of the equatorial ionization anomaly (EIA). Three geomagnetic storms occurred on May 27, 2017 (a month of low occurrence rates of spread-F), October 12, 2016 (a month of transition from low to high occurrence rates of spread-F) and November 7, 2017 (a month of high occurrence rates of spread-F) are analyzed using Global Positioning System (GPS) receivers and ionosondes. The rate of change of total electron content (TEC) Index (ROTI), GPS Ionospheric L-band scintillation, the virtual height of the F-layer bottom side (h'F) and the critical frequency of the F2 layer (foF2) are considered. Furthermore, each ionogram is manually examined for the presence of spread-F signatures.

The results show that, for the three events studied, geomagnetic activity creates favorable conditions for the initiation of ionospheric irregularities, manifested by ionogram spread-F and TEC fluctuation. Post-midnight irregularities may have occurred due to the presence of eastward disturbance dynamo electric fields (DDEF). For the May storm, an eastward over-shielding prompt penetration electric field (PPEF) is also acting. A possibility is that the PPEF is added to the DDEF and produces the uplifting of the F region that helps trigger the irregularities. Finally, during October and November, strong GPS L band scintillation is observed associated with strong range spread-F (SSF), that is, irregularities extending from the bottom-side to the topside of the F region.

---

## 1. Introduction

The response of the low-latitude ionosphere to geomagnetic storms is a prominent topic of study in space weather. There is significant interest in describing the short-term variability of the ionosphere and developing prediction models for ionospheric weather.

The ionospheric irregularities occurrence pattern can be modified drastically during magnetic storms and it may affect the GNSS and VHF signals, so the analysis of the occurrence of irregularities during geomagnetic storms has important applications in navigation and positioning systems, as well as in trans-ionospheric communications.

In this work the ring current Dst index is used to classify the geomagnetic storms (Gonzalez, Tsurutani, & Clúa De Gonzalez, 1999). If the minimum Dst is < -100 nT the storm is intense, -100 nT ≤ Dst < -50 nT corresponds to a moderate storm and -50 nT ≤ Dst < -30 nT characterize a weak storm.

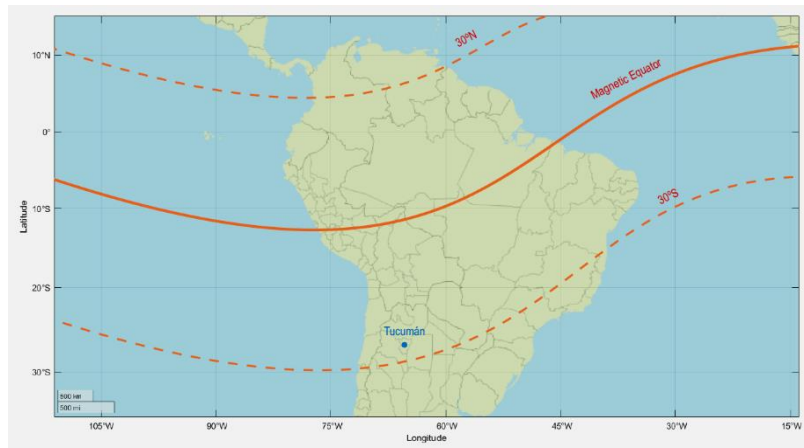

*Figure 1. Map showing the location of Tucumán (blue circle).*

The Dst index is based on the depression in the H component of the geomagnetic field at low latitudes caused by an enhancement in the ring current during the storm.

Low-latitude electric fields can be significantly disturbed during storms. Two main high-latitude sources of these changes are: The solar wind-magnetosphere dynamo and the ionospheric disturbance dynamo. The first one generates rapid and short-lived (2–3 h) (prompt) electric field perturbations associated with rapid changes in the polar cap potential (Senior and Blanc, 1984). When the polar cap potential increases suddenly the situation is called "under-shielding", and the associated electric field (PPEF) has eastward polarity during the day and westward polarity after ~22LT. On the other hand, an over-shielding electric field (PEF) is associated with the recovery of the polar cap potential and has its polarity opposite to that of the PPEF. The ionospheric disturbance dynamo results from thermospheric disturbance winds generated by Joule heating at auroral latitudes during periods of high magnetic activity (Blanc and Richmond, 1980). It produces disturbance wind dynamo electric fields (DDEF) that could last several hours and has a polarity local time dependence that is opposite to that of the PEF. The DDEF is delayed by a few hours with respect to the storm onset (M. a. Abdu, 2012; B. G. Fejer, Larsen, & Farley, 1983; Bela G. Fejer, Jensen, & Su, 2008).

The ionospheric instabilities in F region are grouped under the name of Spread-F. This term was coined to describe the effect of broadening in frequency (frequency spread-F, FSF) and / or in range (range spread-F, RSF) observed in echo traces of ionograms. This is due to multiple reflection paths created by the turbulent ionosphere when there is a process of instability above the ionosonde. Spread-F extends from the F region to 1700 km, and is a nighttime phenomenon. During quiet geomagnetic conditions it occurs mainly before midnight (M A Abdu, de Medeiros, Sobral, & Bittencourt, 1982; Calvert, 1962; Piggott & Rawer, 1972). RSF is associated with the development of plasma bubbles (PB) (Abdu *et al.*, 2003), regions of very low plasma density and a high electric field. These plasma irregularities develop through the Rayleigh-Taylor instability (RTI) process, which operates at the bottom side of the F-region. In the South American longitude sector there are distinct RSF/PB seasons, with high occurrence of RSF during the December solstice months (November to February) and low occurrence during the June solstice months (May to August), while the equinox months (March, April, September and October) presents transition characteristics from high to low occurrence and vice versa.

There has been several attempts to find a correlation between the geomagnetic activity and the occurrence of RSF/PB at low latitudes (M. A. Abdu et al., 2012; Jayachandran, Ram, Somayajulu, & Rao, 1997; Martinis, 2005; Pavlov et al., 2006; Ray, Roy, & Das, 2015, and references therein). Earlier studies show that the RSF is reduced during disturbed geomagnetic conditions (Lyon, Skinner, & Wright, 1960). More recent works conclude that at low latitudes during the low equatorial plasma bubbles occurrence season and transition season, geomagnetic activity helps in the generation process of PB, whereas during the high PB season it acts as inhibitor (Becker-Guedes et al., 2004). Some authors have reported the occurrence of post-sunset RSF during the main phase of a geomagnetic storm for periods of low occurrence rates of spread-F. Basu et al. (2001) analyzed the augmentation or inhibition of spread-F during two major geomagnetic storms for low and middle latitudes. They concluded that PPEF generate post-sunset spread-F at low latitudes and coincides with an increased in AE index and a decrease in SYM-H index. They also showed that the time variation of the SYM-H is an indicator for the time of prompt penetration. Other studies concluded that, depending on the phase of the storm, geomagnetic activity can either suppress or trigger the generation of spread-F in the post-sunset period (pre-midnight). A consensus is that the probability of spread-F occurrence during the post-midnight period increases with geomagnetic activity (Bowman, 1991; Sobral et al., 1997).

This work reports for the first time the influence of three geomagnetic storms on the occurrence of RSF/PB in Tucuman—Argentina. Data from ionosonde and Global Positioning System (GPS) are used. The geomagnetic storms occurred on 2016 and 2017, at the end of the descending phase of the 24th solar cycle. The chosen seasons are winter and summer of 2017 (May and November) and equinox of 2016 (October). Two of these storms, the one occurred in May and the one occurred in October are caused by a coronal mass ejection (CME) whereas the storm occurred in November is caused by a high-speed solar wind stream (HSSWS).

## 2. Data and methodology

The different phases of a geomagnetic storm (initial, main and recovery) are determined by the variation in Dst geomagnetic index. This index is obtained from the World Data Center (WDC) Kyoto, Japan website http://wdc.kugi.kyoto-u.ac.jp/dstae/ index.html. The north-south component of the interplanetary magnetic field (IMF_$B_z$) obtained from the Advanced Composition Explorer (ACE) and the dawn-to-dusk interplanetary electric field (IEF_$E_y$) are analyzed. Additionally, Kp (a 3-hourly planetary index of geomagnetic activity) and AE (a geomagnetic index of the auroral electrojet) are taken from the NASA's Space Physics Data Facility (NASA, Goddard Space Flight Center, https://omniweb.gsfc.nasa.gov). The ionospheric sounding data is obtained by two ionosondes: The Advanced Ionospheric Sounder (AIS) and the Vertical Incidence Pulsed Ionospheric Radar (VIPIR). Both instruments are located in Tucumán (26.9 ° S, 294.6 ° E, lat. Geomagnetic 15.5 S). Figure 1 shows the geographic location of the analyzed region. The sweeping frequency of the AIS ionosonde is from 1 to 20 MHz and the sounding repetition rate is 10 minutes, the ionograms are available at the electronic Space Weather upper atmosphere database (eSWua) (http://www.eswua.ingv.it/). The VIPIR operates between 0.3 and 25 MHz with a sounding repetition rate of 5 minutes (Bullett, 2008), the ionograms can be obtained from the website of the Low Latitude Ionospheric Sensor Network (LISN)

(http://lisn.igp.gob.pe). Each ionogram is manually examined for the presence of RSF. Also, the virtual height of the F-layer bottom side, h'F and the critical frequency of the F2-layer, foF2, are extracted. For the AIS ionosonde, the parameters are auto scaled by the Autoscala system (Pezzopane and Scotto 2005) and for the VIPIR ionosonde, the parameters are manually scaled.

The Total Electron Content (TEC) is obtained from a GPS ground-based receiver located at Tucuman, and the raw GPS observables are available at the Argentine Continuous Satellite Monitoring Network (RAMSAC) website (http://www.ign.gob.ar/NuestrasActividades/Geodesia/Ramsac) (Piñón *et al.*, 2018). The slant TEC along the satellite-receiver line of sight is estimated with the GPS-TEC calibration technique developed by Dr. Krishna Seemala Gopi of the Indian Institute of Geomagnetism (IIG), Navi Mumbai, India (GPS_Gopi_v2.9.5:). An elevation mask of 25° has been applied to reduce the effects of multipath.

Ionospheric storms have been categorized by as positive and negative phases. A positive phase results in increased electron density from the quiet time values. Whereas a negative phase results in decreased electron density from the quiet time values. The response of the ionospheric F-region to a geomagnetic storm is analyzed using ΔTEC, that is the deviations of TEC from the reference, expressed in Eq. 1. Therefore, ΔTEC > 0 indicates a positive phase and ΔTEC < 0 indicates a negative phase.

$$\Delta TEC = \frac{TEC - \langle \overline{TEC} \rangle}{\langle \overline{TEC} \rangle} \times 100 \qquad (1)$$

$\overline{TEC}$ is the mean TEC considering all the visible satellites during a day, $\langle \overline{TEC} \rangle$ is the average $\overline{TEC}$ calculated using the ten International Quietest Days (IQDs) of the month. IQDs are derived from GFZ-Potsdam (https://www.gfz-potsdam.de/en/kp-index/).

Ionospheric L-band scintillation is also obtained from the GPS receiver. Scintillation index S4 is used to quantified the strength of the amplitude scintillation. S4 > 0.5 indicates strong scintillation and 0.1 < S4 ≤ 0.5 indicates weak scintillation activity (Davies, 1990). The S4 data is obtained from the LISN database for May and October and from the GPS Ionospheric Scintillation receiver owned by Istituto Nazionale di Geofisica e Vulcanologia (INGV) for November.

Furthermore, the rate of change of TEC (ROT) along the signal path from each visible satellite to the receiver and the rate of TEC index (ROTI) are computed. The ROT and ROTI can be used to detect the presence of GPS ionospheric irregularities during magnetic storms (Pi et al., 1997; Basu et al., 1999; Azzouzi et al., 2016; Dugassa, Habarulema, & Nigussie, 2020). The ROT is the rate of change of slant TEC (Eq. 2) and the ROTI is defined as the standard deviation of ROT (Eq. 3).

$$ROT = \frac{TEC_k^i - TEC_{k-1}^i}{t_k - t_{k-1}} \qquad (2)$$

$$ROTI = \sqrt{\langle ROT^2 \rangle - \langle ROT \rangle^2} \qquad (3)$$

Where k is the time of epoch and *i* is the visible satellite. In the present work, the sampling interval used to calculate ROT is 0.5 min and the time window of the standard deviation of ROTI is 5 min.

ROTI is divided into different levels, ROTI < 0.25 indicates no TEC fluctuations, 0.25 ≤ ROTI < 0.5 are considered weak TEC fluctuations, 0.5 ≤ ROTI < 1 signifies moderate TEC fluctuations and ROTI ≥ 1 are strong TEC fluctuations (Atıcı & Sağır, 2019; Liu, Yuan, Tan, & Li, 2016; Ma & Maruyama, 2006).

**3. Results**

*3.1. The Storm of May 27, 2017*

The storm occurred in May, a month of low occurrence rates of spread – F, with a monthly mean F10.7 flux of 76.3 solar flux unit (sfu). The storm reaches a minimum Dst of -125 nT, and a Kp of 7, which makes it is an intense storm. It is produced by a CME released by the sun on May 23, 2017, that arrived at earth four days later. The sudden storm commencement (SSC) occurs at 15:34 UT (12:34 LT) on May 27 (http://www.obsebre.es/en/rapid), the initial phase lasts until 19 UT (16 LT) and the main phase is maintained until May 28 at 7 UT (4 LT) when the Dst index reaches the minimum. Figure 2 shows the IMF_Bz obtained from ACE and the interplanetary electric field IEF_Ey as a function of the universal time UT (LT = UT − 3). The geomagnetic indices Dst, Kp and AE, and the ionospheric parameters h´F and foF2 scaled from the ionograms are also plotted in the same figure. The periods with RSF are highlighted with vertical bars.

It is observed that Bz presents a strong southward excursion during the storm main phase, between 20 UT (17 LT) on May 27 and 15 UT (12 LT) on May 28, with a minimum of -19.5nT at 0 UT on the same day (21 LT May 27), then it turns northward and remain thus during about 19 hours with a peak of 11.6 nT at 22 UT on May 28. The interconnection between the IMF and the Earth´s magnetic field is produced during the southward Bz period and results in the large decrease observed in Dst. $E_y$ shows a large increase during the main phase and has a maximum of 7.72 mV/m at 23 UT (20 LT) on May 27. The AE index (which is a measure of currents in the auroral electrojet) shows a small peak of 361 nT at 16 UT (13 LT) on May 27, during the initial phase of the storm. It then increases from 34 nT to 943 nT between 20 – 23 UT (17 – 20 LT). Small-amplitude fluctuation can be seen at 23 – 5 UT (20 – 2 LT) on May 28 (storm main phase) with a peak of 1271nT. Finally, AE decreases to quiet values during the recovery phase of the storm. The foF2 data shows an increase of 1.7MHz during the initial phase of the storm compared to the quiet-time levels (overage of the 10 IQDs of May 2017) with a peak of 8.3 MHz at 18 UT (15 LT) on May 27. A larger intensification of about 4.5 MHz occurs during the recovery phase, with a peak of 10.3 MHz at 16 UT (13 LT) on May 28. The bottom panel shows the variation of h´F, it is observed that during the initial phase h´F is close to its quiet-time reference value, then on May 28, h´F increases from ~213 km to 357 km at 0 – 4 UT (21 – 1 LT), it increased 54% more than the quiet time curve in the same period. This behavior coincides with the large sudden increase in the AE index during the second half of the main phase. Finally, during the recovery phase, h'F decreases irregularly to quiet levels.

To study the Earth´s electric field penetration, ΔH is used to infer the electric field at low latitudes. ΔH is the difference in the magnitudes of the horizontal geomagnetic field component (H) between a magnetometer placed on the magnetic equator and one displaced 6°- 9° away. As it is explained by Wei et al. (2015), ΔH is related to the equatorial electrojet (EEJ) and the EEJ is linearly related to the electric field. Figure 3 shows the difference between H at Jicamarca (dip latitude 0.4°N) and Piura (dip latitude 6.8°N), Peru. A weak positive perturbation of ΔH is observed on May 28 at 3 – 8 UT (0 – 5 LT), that corresponds to an eastward electric field likely associated to Bz slowly turning north.

Figure 4 shows the day-to-day variability of RSF over Tucuman during May 2017, observed with the ionosonde AIS. The y-axis represents the day of the month and the x-axis represents the hour (UT) of the day. The graphic shows that RSF is present in four days: the three most disturbed days of May (28, 20 and 19) and one quiet day (May 5) at 1 – 4:30 LT. During the period of the storm, the ionograms show RSF during the second part of the main phase, in the interval 01:40 UT – 07:40 UT (May 27 22:40 LT – May 28 04:40 LT). During the disturbed period, RSF appears before local midnight and it is more intense. To see the evolution of the spread-F on May 28 at Tucuman, figure 5 illustrates the beginning, evolution process, and end of the irregularities.

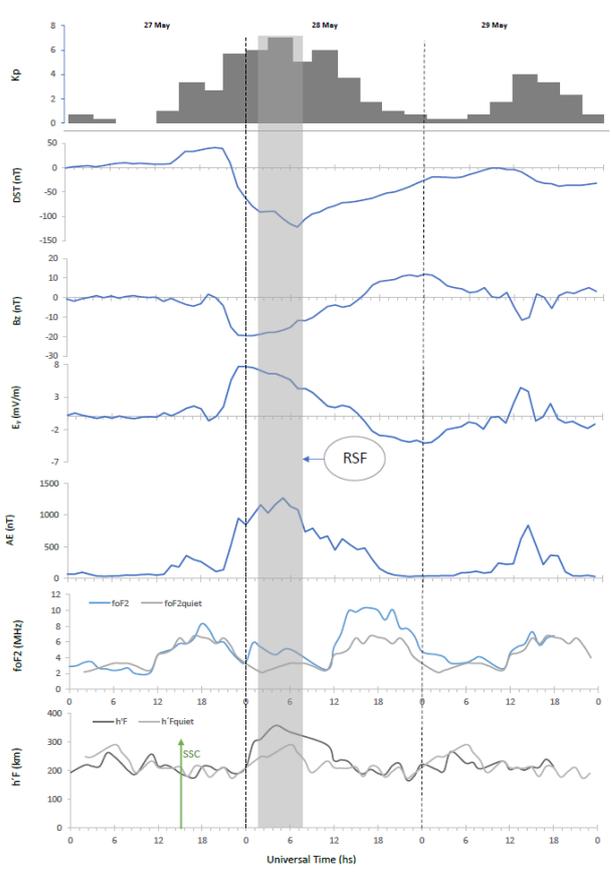

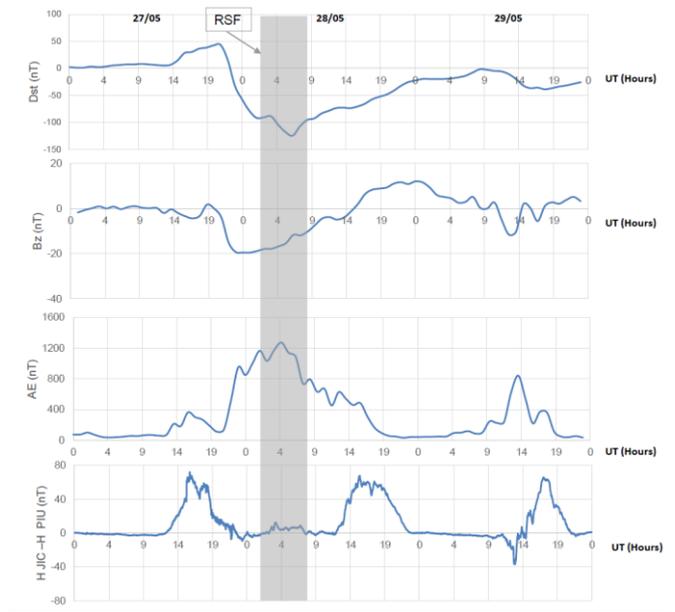

Figure 3. Time variations of Dst (nT), Bz (nT), AE (nT) and difference between horizontal geomagnetic field components (H) at Jicamarca and Piura, Peru, during May 27 – 29, 2017. The shaded region indicates the occurrence of RSF.

Figure 2. Kp index, Dst, Bz, Ey, AE, foF2 and h´F for Tucumán during May 27 – 29, 2017. The shaded region indicates the occurrence of RSF.

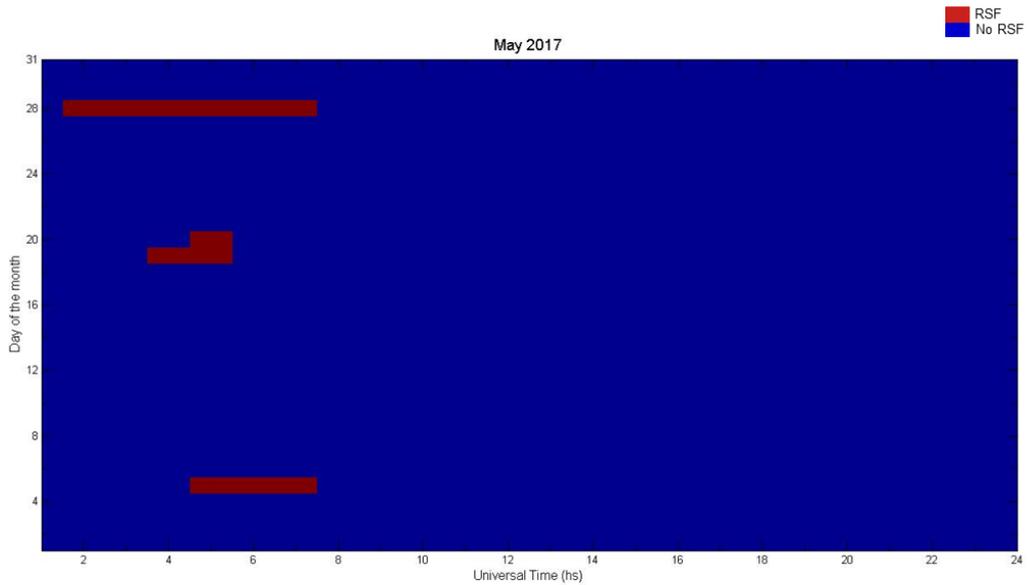

Figure 4. Day-to-day variability of RSF occurrence over Tucuman during May 2017, RSF (red).

Figure 6 shows the TEC estimated from a GPS receiver at Tucuman from different satellites, it is possible to see TEC depletions on May 28 at ~1 – 7 UT (22 – 4 LT) for most of the satellites in view. In order to analyze this data segment more deeply, the TEC perturbations (TECp) along PRNs 12 and 15 arcs are calculated according to Eq. (4).

$$TEC_p^i(t) = TEC^i(t) - \langle TEC^i(t) \rangle \qquad (4)$$

Where $TEC^i(t)$ is the TEC value along the satellite *i* and the receiver at a time t and $\langle TEC^i(t) \rangle$ is the corresponding 1 h running mean.

Fast Fourier Transform (FFT) analysis (periodogram) is performed on the data to identify different periods. Figure 7 shows the FFT of TEC perturbations characterizing PRNs 12 and 15 on May 28 2017 from 1 UT to 7 UT. It is observed that dominant periods of ~70 and ~40 minutes are present in both cases. These TEC depletions could be due to the propagation of Atmospheric Gravity Waves (AGW) in the ionospheric F region that generate traveling ionospheric disturbances (TIDs) (Hines, 1959; Hooke, 1968; Hunsucker, 1982; Kirchengast et al., 1996; Valladares et al., 2009).

Figure 8a and b shows variations of VTEC, ROT index and ROTI index over Tucumán for PRN 12 and 15 during May 27-29, 2017. As it was mentioned before, TEC profile for both PRNs is characterized by depletions on May 28, while it is smooth on May 27 and 29. On May 28 between 1 – 4 UT, for PRN 12 and 15, ROT level is ~ 1 TECU/min and ROTI presents values of 0.5 – 1 TECU/min, this indicates moderate TEC fluctuations.

The analysis of the scintillation index S4 (figure 9) for the period of the storms reveals a weak scintillation activity given that S4 is always between 0.1 and 0.3. Thus, the electron density irregularities associated with the PB observed in the period of the storm do not cause significant amplitude scintillation.

Figure 10 shows the deviation, $\Delta TEC$, for May 27 – 29, 2017 calculated with Eq. (1). The data presents an irregular behavior with small effects during the initial and the first part of the main phase; first a negative ionospheric storm phase is present on May 27 between 15:30 UT and 17:30UT (12:30 LT – 14:30 LT) followed by a short-lived positive phase between 17:30 UT – 19 UT (14:30 LT – 16 LT) and then a negative phase again from May 27 at 19 UT (16 LT) to May 28 at 23:30 UT (20:30 LT). During the last part of the main phase, a positive ionospheric storm effect is observed on May 28 between 23:30 UT and 6 UT (20:30 LT – 3 LT) with a peak of 96% at 2 UT on May 28 (May 27, 23 LT) followed by minor negative disturbances during the first part of the recovery phase on May 28 between 6 UT and 10 UT (3 LT – 7 LT). A positive storm is observed on May 28 between 10 UT (7 LT) and 22 UT (19 LT), positive values of ΔTEC exceed 100% and the peak enhancement occurs almost 7 hours after the minimum Dst. Finally, an irregular behavior is observed during the recovery phase, with minor positive and negative disturbances.

### 3.2. The Storm of October 12, 2016

A CME that hit our planet on October 12 (a month of transition from low to high occurrence rates of spread-F) at 22:12 UT (19:12 LT) caused a geomagnetic storm with a minimum Dst of -104 nT. For this month the mean F10.7 index is 84.6 sfu. Figure 11 shows the geomagnetic indices and the F-layer parameters foF2 and h'F during October 12 – 14, 2016. The main phase of the storm starts at 6 UT (3 LT) on October 13

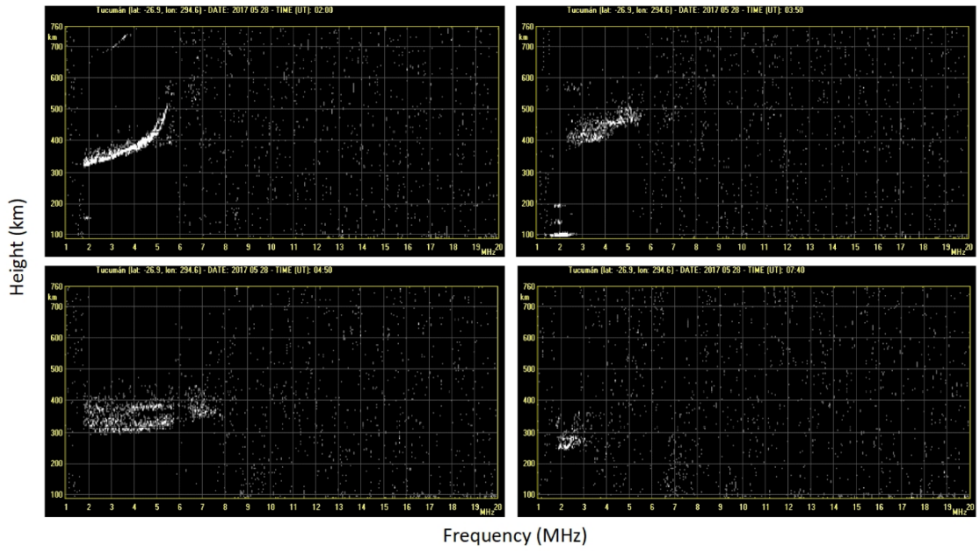

*Figure 5. Ionograms recorded at Tucuman showing the presence of RSF on May 28, 2017.*

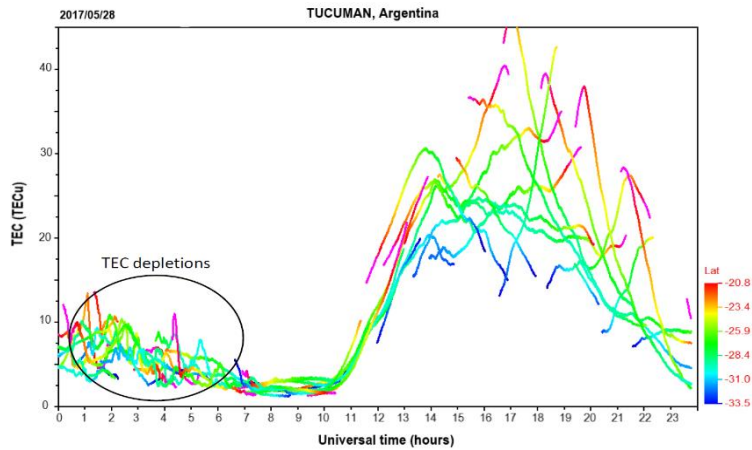

*Figure 6. TEC calibrated from a GPS receiver at Tucuman. TEC depletions are indicated with a black circle on May 28 at ~1 – 7 UT (22 – 4 LT).*

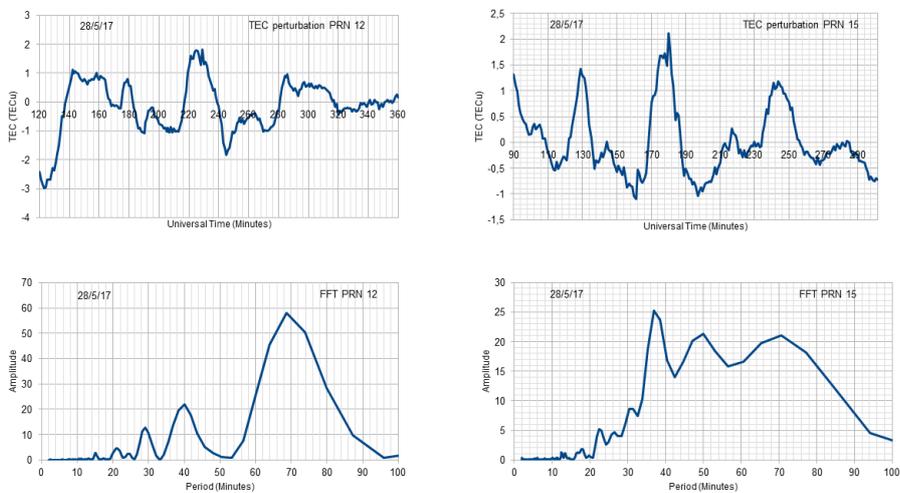

*Figure 7. Upper panel, the TEC perturbations characterizing PRNs 12 (left) and 15 (right) on May 28, 2017. Bottom panel, FFT of TEC perturbations characterizing PRNs 12 (left) and 15 (right).*

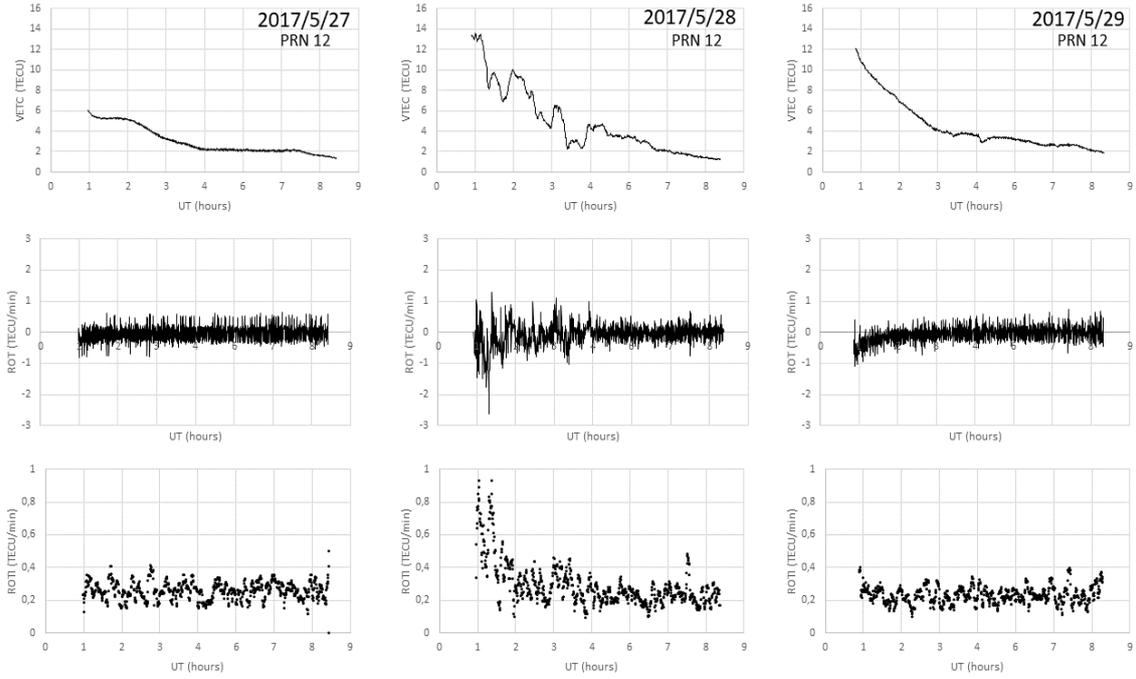

*Figure 8a. Variations of VTEC, ROT and ROTI for PRN 12 over Tucumán during May 27 – 29, 2017.*

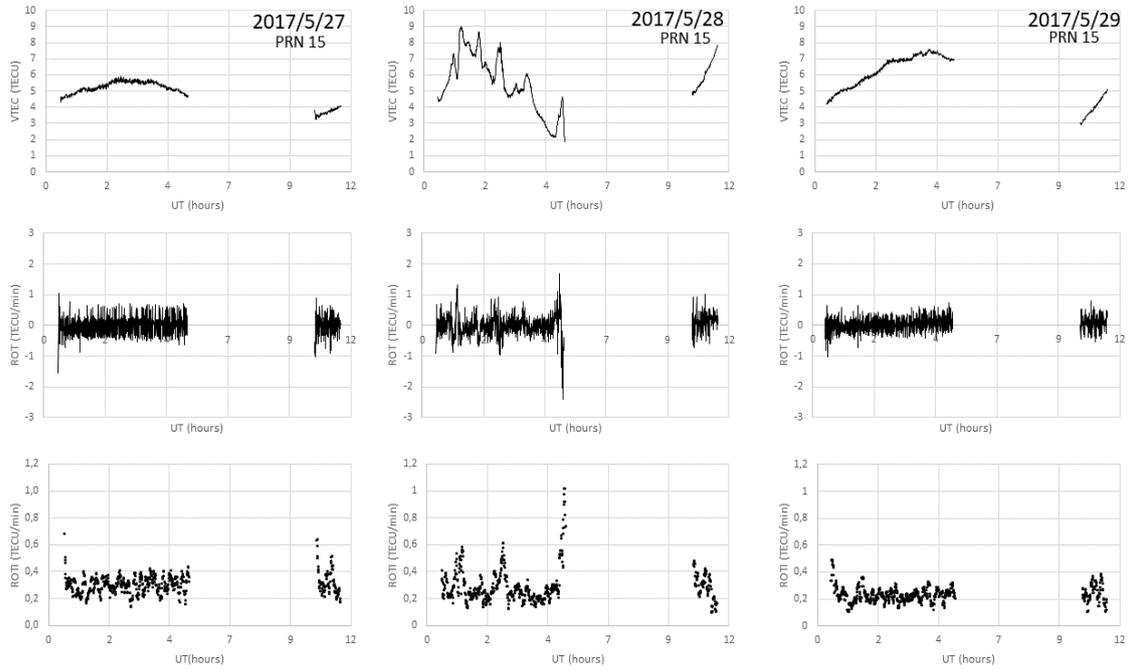

*Figure 8b. Variations of VTEC, ROT and ROTI for PRN 15 over Tucumán during May 27 – 29, 2017.*

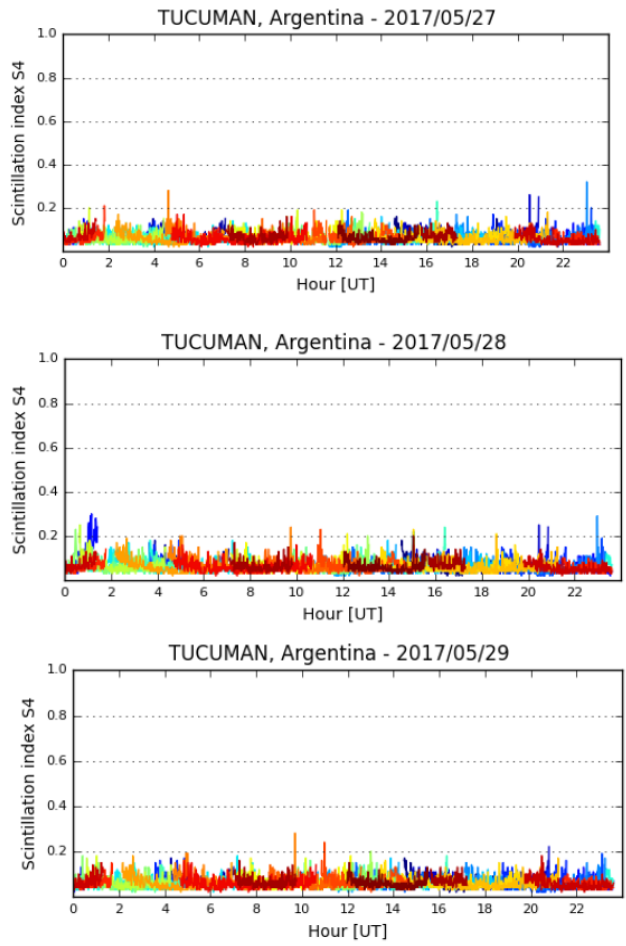

*Figure 9. The temporal variation of Global Positioning System L-band scintillation over Tucuman on May 27, 28 and 29 2017.*

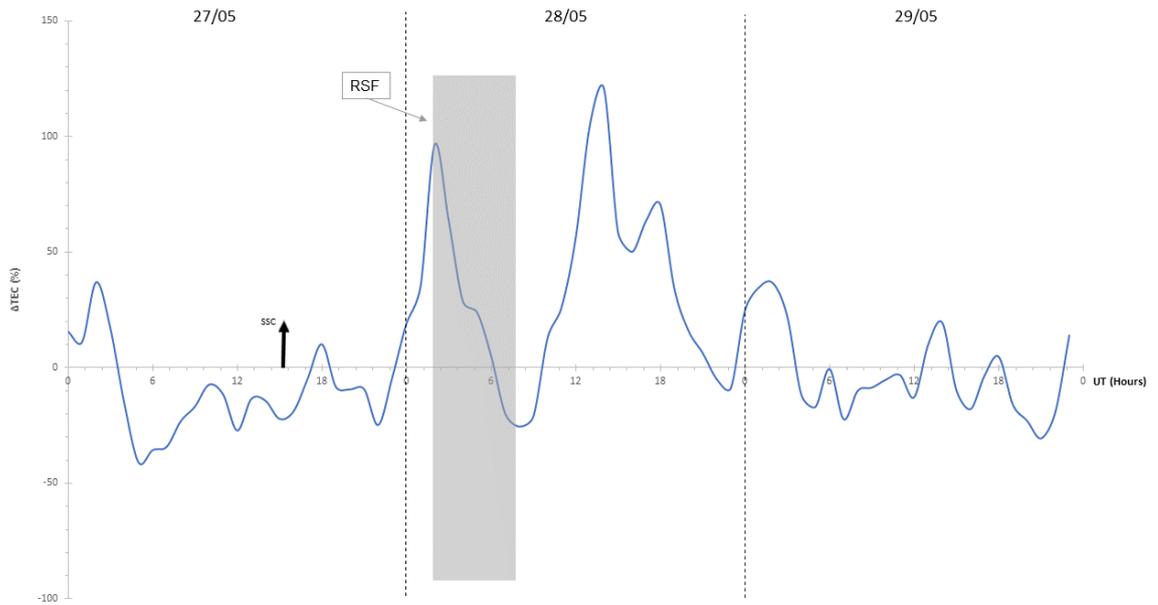

*Figure 10. Deviation ΔTEC between the TEC values for May 27- 29, 2017 and the average TEC of the 10 quietest days of May 2017.The shaded region indicates the period when spread-F is observed in ionograms.*

and remains until 17 UT (14 LT) followed by a gradual recovery. The highest Kp is 6 and occurs at 15 UT (12 LT) on October 13. AE and $E_y$ show a rapid increase coinciding with an intense Bz south condition from October 13 at 6 UT (3 LT) till October 14 at 9 UT (6 LT), Bz has a minimum value of -20.8 nT at 16 UT (13 LT) on October 13, AE has a maximum value of 1200 nT at 15 UT (12 LT) while the highest value of $E_{sw}$ is 16 mV/m at 16 UT (13 LT). During the recovery phase of the storm, Bz turns north while AE and E decrease to quiet values.

It is observed that during the period of the storm, foF2 is generally higher than the quiet values (overage of the 10 IQDs of October 2016), specially between 18 UT (15 LT) and 6 UT (3 LT). The largest difference is 4.5 MHz and occurs at 2 UT on October 13 (23 LT October 12), during the initial phase of the storm. The peak value of foF2 is 16.6 MHz at 22 UT (19 LT) on October 13.

As for h'F, it is observed that the disturbed values are usually higher than the quiet ones, except during the initial phase when the values for both periods are similar. During the recovery phase on October 14, h'F increases from ~216 km to ~336 km at 1 – 4 UT (22 – 1 LT), ~38% more than the quiet value and during the main phase, on October 13 around 15 UT (12 LT) h'F is ~22% higher than for quiet days. There is no data for h´F and foF2 for the periods with RSF as it is observed in the curves of figure 7.

Time variations of ΔH for October 12-14 are shown in figure 12. A negative perturbation is observed on October 13 at 9-13 UT (6-10 LT) that can be associated with a westward electric field. ΔH does not present perturbations during the periods with ionospheric irregularities.

Figure 13 illustrates the day-to-day variability of RSF over Tucumán during October 2016. RSF is present in six days: three of the most disturbed days of the month (25, 13 and 29) and one of the ten quietest days (October 19) between 2 – 6 UT (23 – 3 LT), except on October 14 and October 29 when RSF is also observed after 6 UT. During the period of the storm, spread-F is observed during the initial phase, on October 13 at 3:30 – 4:20 UT (0:30 – 1:20 LT), and during the recovery phase on October 14 at 8:50 – 11:50 UT (5:50 – 8:50 LT), indicated with gray bars in figure 12. Two of these ionograms are shown in figure 14. It is observed that in these periods the range spread on F layer echo extends to higher frequencies (~15 MHz), beyond the local foF2 value (~11 MHz), than that present during the storm of May 27 (~8 MHz). This type of spread-F is often called Strong range Spread-F (SSF) and considered as an independent type of spread-F.

TEC depletions are observed on October 13 and 14 in coincidence with the presence of RSF in ionograms as is shown in figure 15. In the same way as in the previous storm, periodogram analysis is performed in the TECp to identify different periods. PRN 1 and 27 are considered, the periods found are ~ 48 and 34 minutes (see figure 16).

On October 13 between 2-3 UT, VTEC depletions are present in PRN 27, ROT levels are ~ 1 TECU/min and ROTI values are 0.6 – 0.8 TECU/min (fig. 17a). On October 14, PRN 1 shows TEC depletions between 9-10 UT, ROT levels are ~3 TECU/min and ROTI is 1.2 – 1.8 TECU/min, this indicates strong TEC fluctuations (fig. 17b).

Figure 18 shows the temporal variation of the scintillation index S4 over Tucuman on October 13 and 14, 2016. Strong scintillation (i.e., S4 ≥ 0.5) is observed on October 13 at 1 – 5 UT (22 – 3 LT) and on October 14 around 10 UT (7 LT). Ionograms SSF and GPS – TEC fluctuations occur almost simultaneously with high amplitudes of S4. Likely the strong scintillation activity could be associated with the field-aligned irregularities (FAIs) with a spatial scale of a few hundred meters that are confined

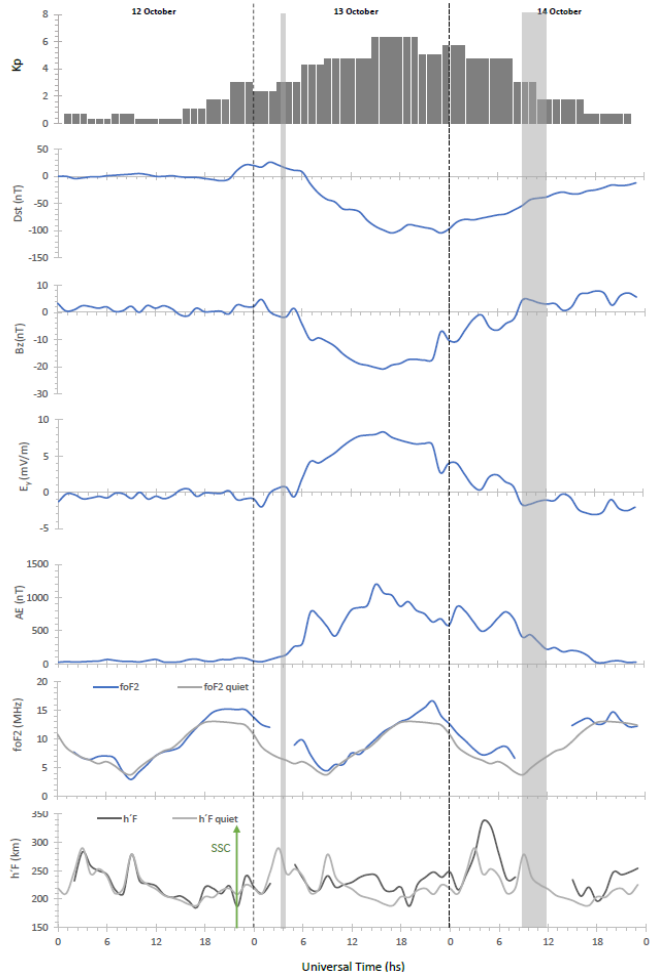

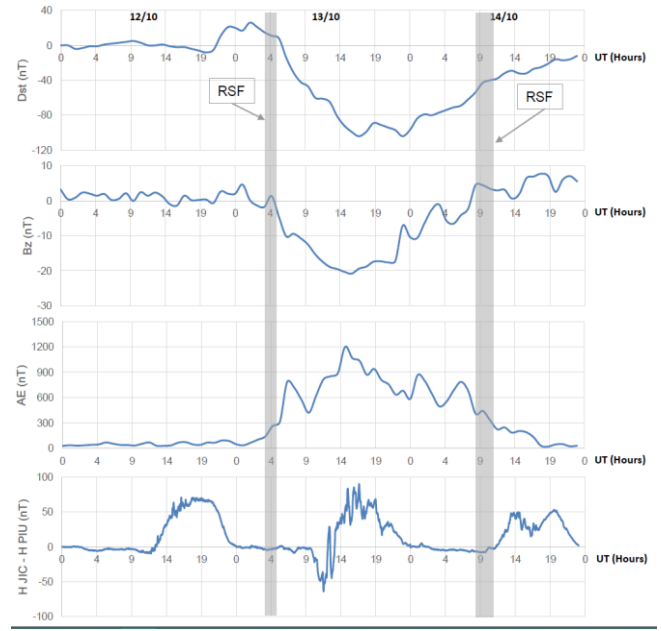

Figure 12. Time variations of Dst (nT), Bz (nT), AE (nT) and difference between horizontal geomagnetic field components (H) at Jicamarca and Piura, Peru, during October 12 – 14, 2016. The shaded regions denote the periods with spread-F.

Figure 11. Kp index, Dst, Bz, Ey, AE, foF2 and h´F for Tucumán during October 12 – 14, 2016. The shaded regions denote the periods with spread-F.

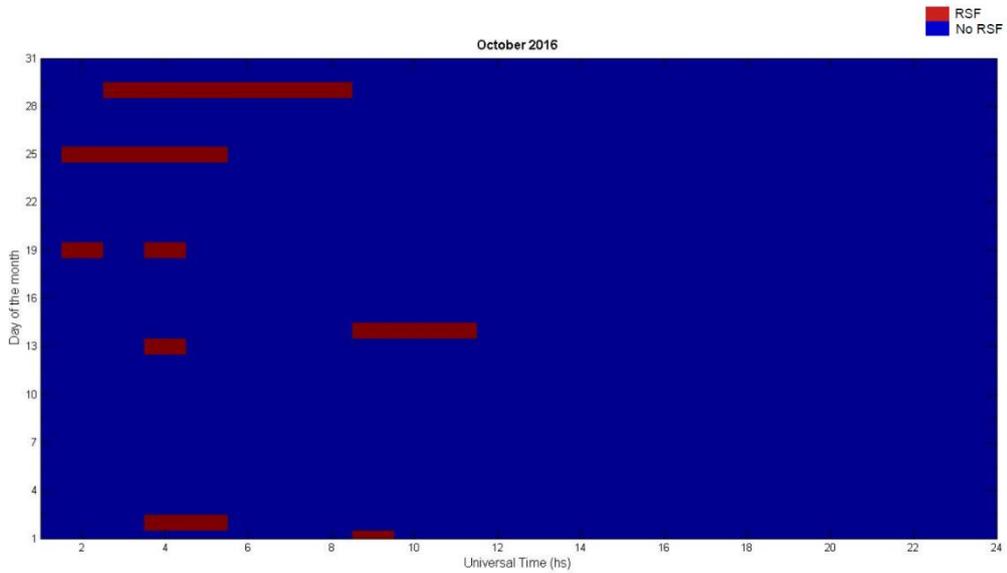

Figure 13. Day-to-day variability of RSF occurrence over Tucuman during October 2016, RSF (red).

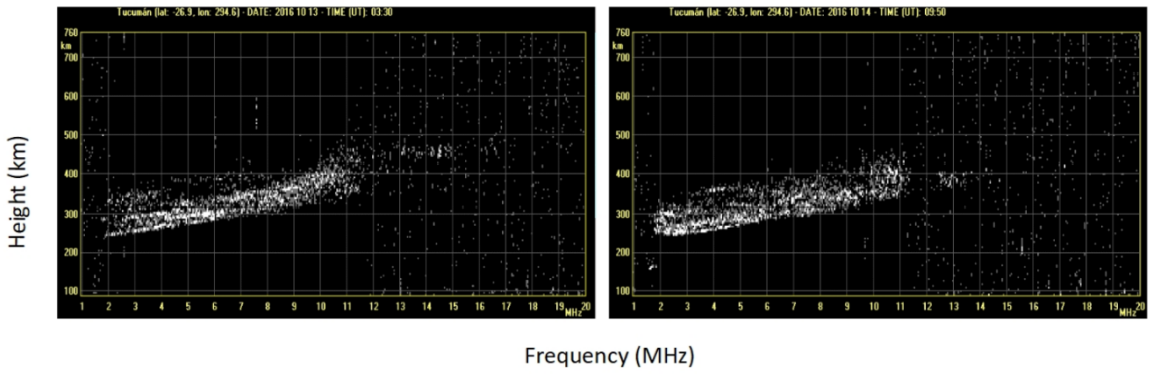

*Figure 14. Sample of strong spread-F (SSF) observed in ionograms recorded using the AIS ionosonde at Tucuman – Argentina on October 13 (left) and 14 (right), 2016.*

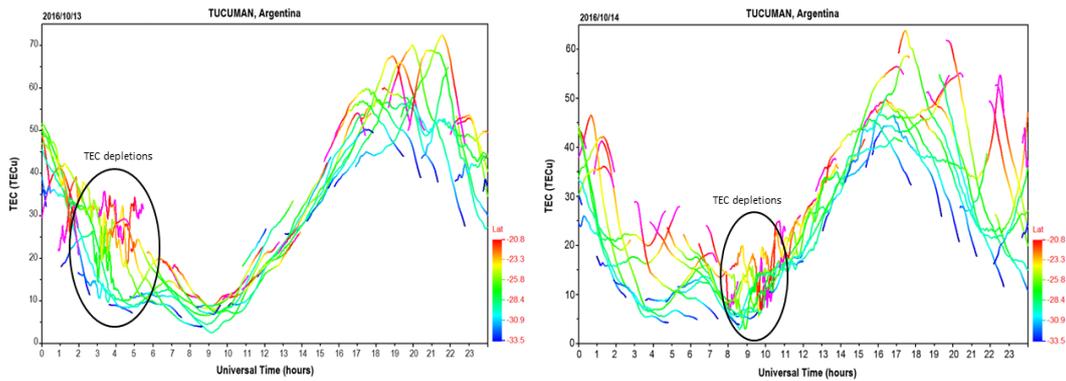

*Figure 15. TEC calibrated from a GPS receiver at Tucuman. TEC depletions (indicated with a black circle) are observed on October 13 at ~3 – 5 UT (0 – 2 LT) and on October 14 at ~8 – 11 UT (5 – 8 LT).*

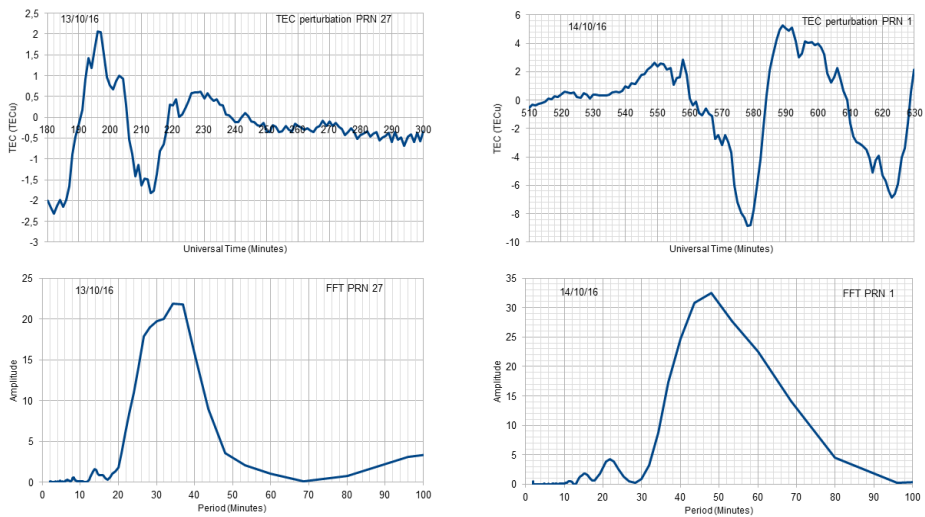

*Figure 16. Upper panel, the TEC perturbations characterizing PRN 27 on October 13, 2016 (left) and PRN 1 on October 14 (right). Bottom panel, FFT of TEC perturbations characterizing PRN 27 (left) and PRN 1 (right).*

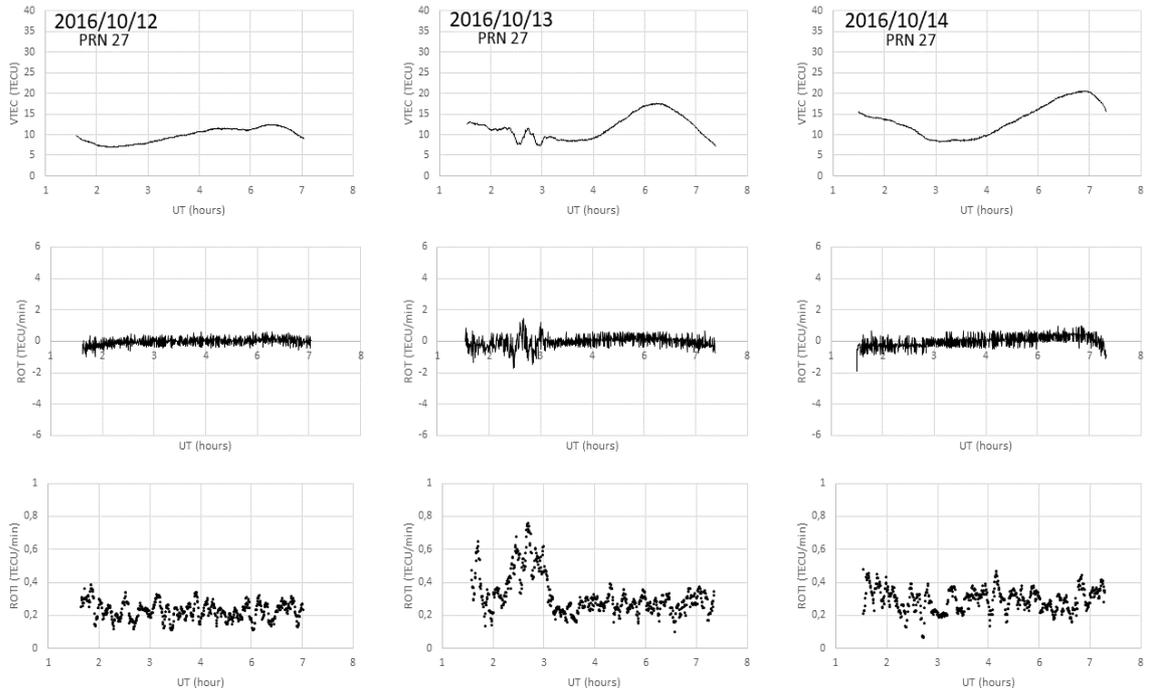

*Figure 17a. Variations of VTEC, ROT and ROTI for PRN 27 over Tucumán during October 12 – 14, 2016.*

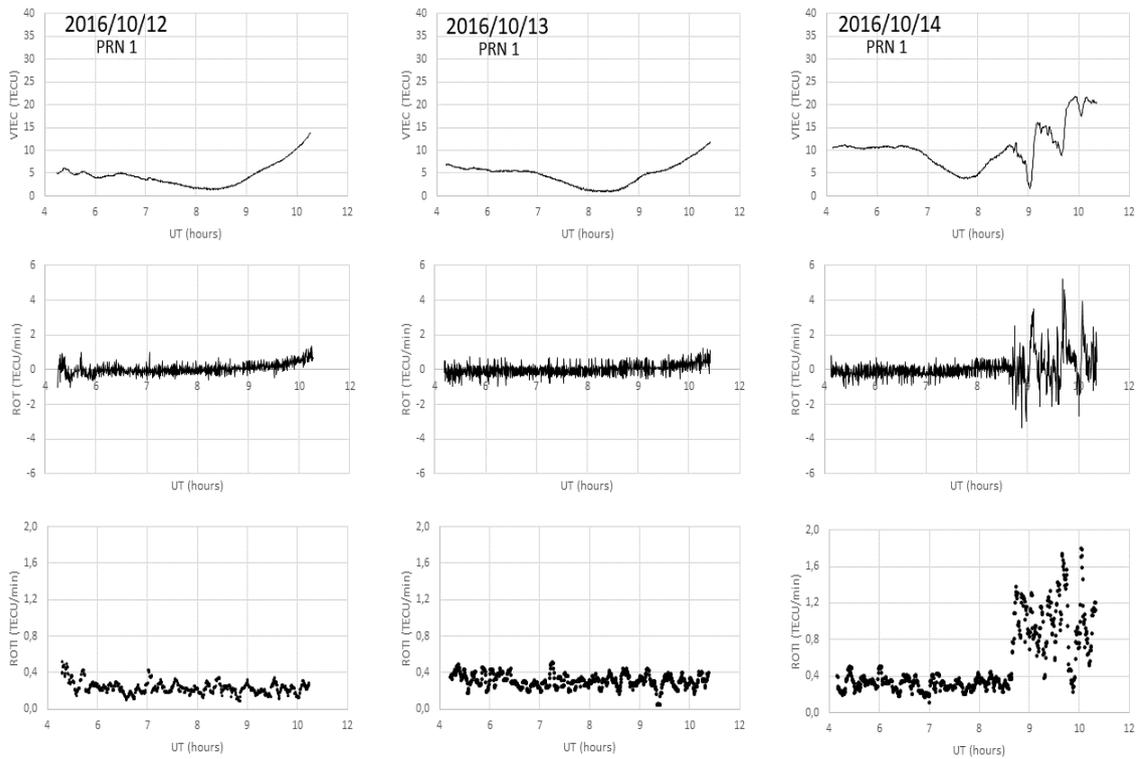

*Figure 17b. Variations of VTEC, ROT and ROTI for PRN 1 over Tucumán during October 12 – 14, 2016.*

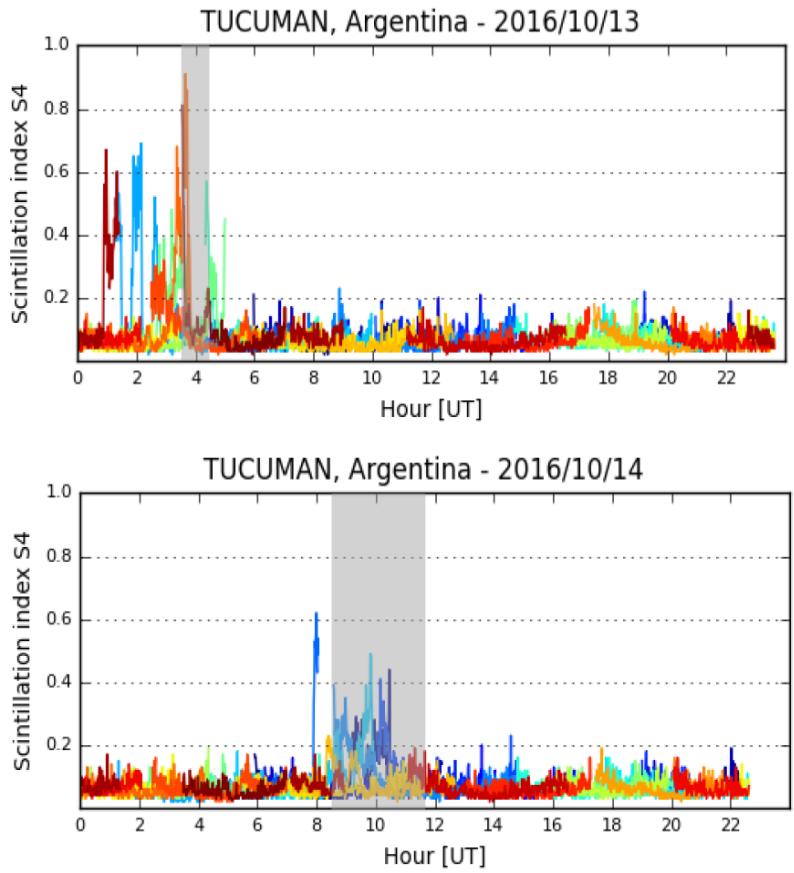

*Figure 18. The temporal variation of Global Positioning System L-band scintillations over Tucuman on October 13 and 14, 2016. The shaded region indicates the periods when SSF is observed in ionograms*

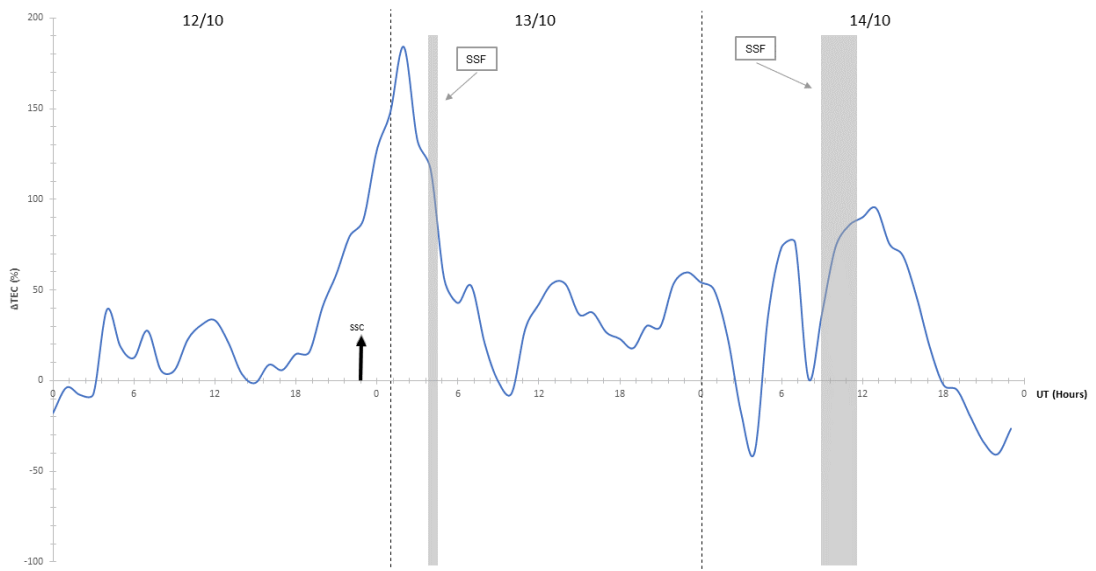

*Figure 19. Deviation ΔTEC between the TEC values for October 12 – 14, 2016 and the average TEC of the 10 quietest days of October 2016. The shaded region indicates the periods when SSF is observed in ionograms.*

within the PBs (Otsuka, Shiokawa and Ogawa, 2006).

$\Delta TEC$ vs UT is shown in figure 19 for October 12 – 14. The dominant feature is the absence of significant negative disturbances and the presence of a large positive effect during the initial phase of the storm with a peak of 184% at 2 UT on October 13 (23 LT October 12). On October 14, 3 – 5 UT (0 – 2 LT) a negative effect is observed, with a peak of -39.7% at 4 UT (1 LT) on October 14.

### 3.3. The Storm of November 7, 2017

A moderate geomagnetic storm occurs on November 7, 2017 (a month of high occurrence rates of spread-F) caused by the impact of high-speed solar wind streams (HSSWS) emanated from a solar coronal hole, with speeds near 650 km/s. The fast streams interact with the slow streams forming and interface region known as Co rotating Interaction Region (CIR). An important aspect of the CIRs is the presence of Alfvén waves in the magnetic field.

During this month the mean F10.7 is 70.3 sfu and the Kp reaches a maximum of 6.3 on November 7 at 18 UT (15 LT) and on November 8 at 3 UT (0 LT). A gradual initial phase (not a sudden commencement) starts on November 7 at ~1 UT (November 6, 22 LT) and finished at about 8 UT (5 LT). The main phase lasts until November 8 at 1 UT (November 7, 22 LT) when the Dst reaches its minimum value of -72 nT. After that, a long recovery phase is observed, IMF Bz oscillations diminish but intense auroral activity is still present with AE values higher than 1000 nT on November 8. This is not a High-Intensity, Long-Duration, Continuous AE Activity, or HILDCAA event since the active conditions last less than two days (Tsurutani & Gonzalez, 1987).

The response of the F-region over Tucuman during this geomagnetic storm is presented in Figure 20. During the period of analysis IMF Bz is highly variable, it oscillates rapidly between north and south. This is in contrast to the CME–driven storms analyzed before that present long-lasting southward and northward incursions. Therefore, the energy injection processes from the solar wind to the magnetosphere-ionosphere system are different. In the CME-driven storms the energy is transferred in large amounts whereas in the CIR-driven storms the energy is transferred by little impulses (Rodríguez-Zuluaga et al., 2016; Tsurutani et al., 2006). In spite of this, during CIR storms the total amount of energy injected into the magnetosphere can be large because they last longer than CME storms. For the event analyzed here, the maximum Bz south is 11 nT and occurs on November 7 at 9 UT (6 LT), Bz remains south between November 7 at 17 UT (14 LT) and November 8 at 0 UT (November 7, 21 LT), then it turns northward and go back southward on November 8 between 1 – 6 UT (22 – 3 LT).

$E_y$ has an irregular behavior mainly with positive values, three peaks are observed: 4.4 mV/m on November 7 at 9 UT (6 LT), 4.3 mV/m at 21 UT (18 LT), 3.8 mV/m at 18 UT (15 LT) and 3.9 mV/m on November 8 at 12 UT (9 LT). AE shows an oscillatory behavior, it tends to increase between 0 UT (21 LT) on November 7 and 6 UT (3 LT) on November 8 and to decrease between 6 UT (3 LT) on November 8 and 6 UT on November 9. The peaks values are 1059 nT at 4 UT (1 LT) and 1070 nT at 12 UT (9 LT) on November 8. The oscillatory behavior in Bz, Ey and AE is associated to the Alfvén waves within the CIR.

As for foF2, it presents similar values that during quiet days except on November 8 between 0 and 15 UT (November 7 21 LT – November 8 12 LT) when the values are ~15% higher than the quiet ones. Regarding h´F, the values during the period of the storm are similar to those for quiet time.

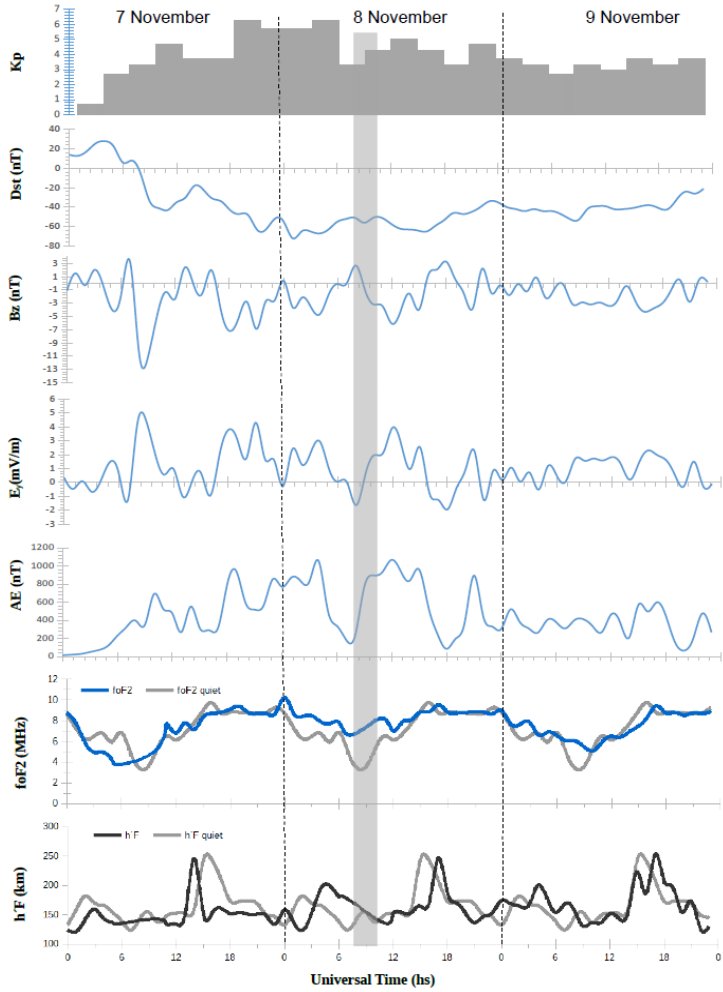

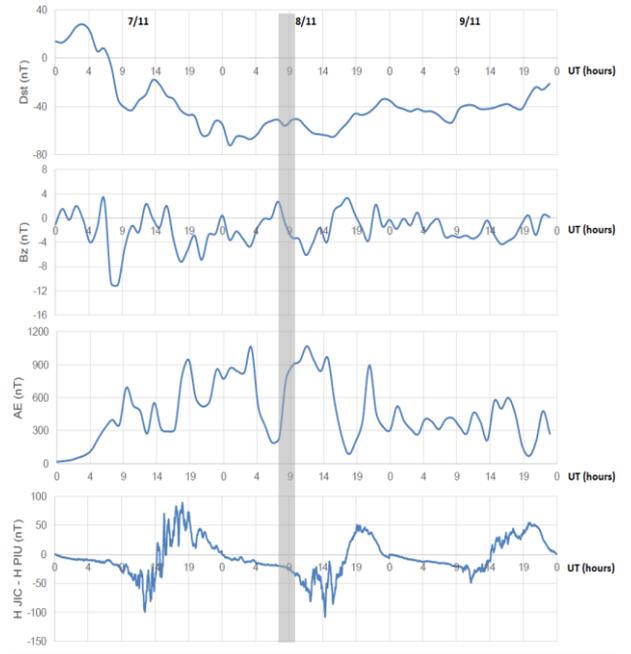

*Figure 21. Time variations of Dst (nT), Bz (nT), AE (nT) and difference between horizontal geomagnetic field components (H) at Jicamarca and Piura, Peru, during November 7 – 9, 2017. The shaded regions denote the periods with spread-F.*

*Figure 20. Kp index, Dst, Bz, Ey, AE, foF2 and h´F for Tucumán during November 7 – 9, 2017. The shaded regions indicate periods with SSF.*

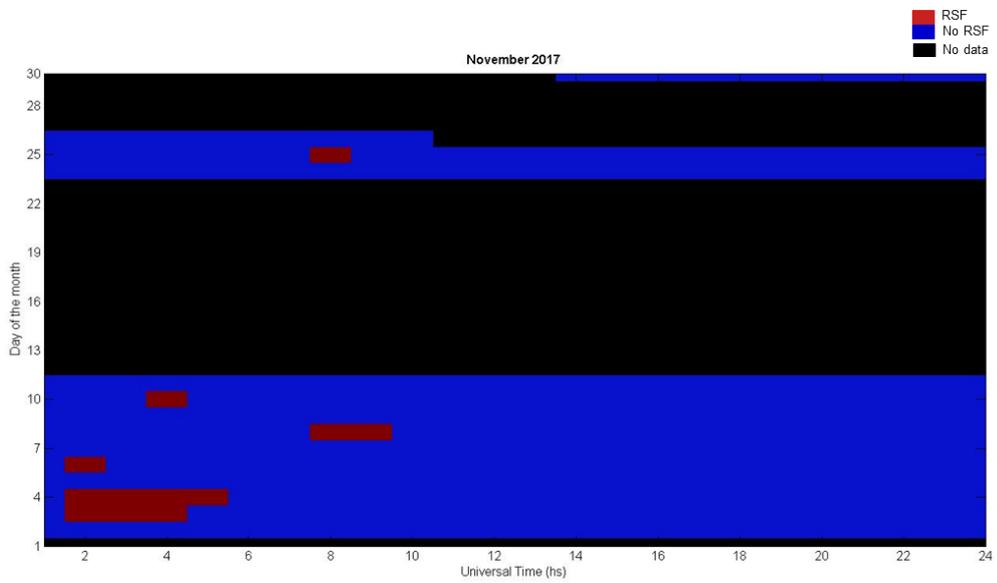

*Figure 22. Day-to-day variability of RSF occurrence over Tucuman during November 2017, RSF (red), data unavailability (black)*

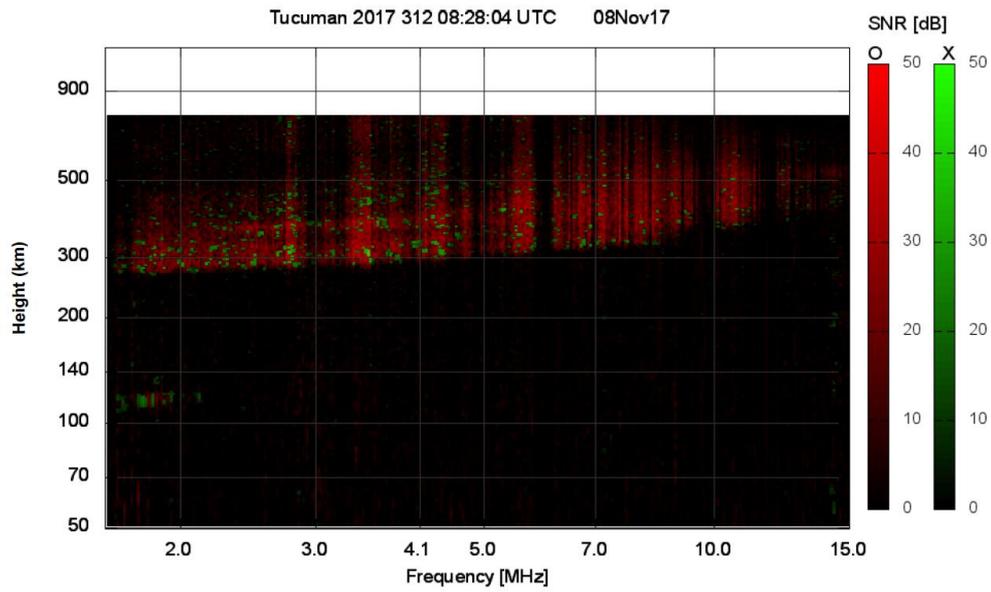

*Figure 23. Sample of strong spread-F (SSF) recorded using the VIPIR ionosonde at Tucuman – Argentina on November 8, 2017.*

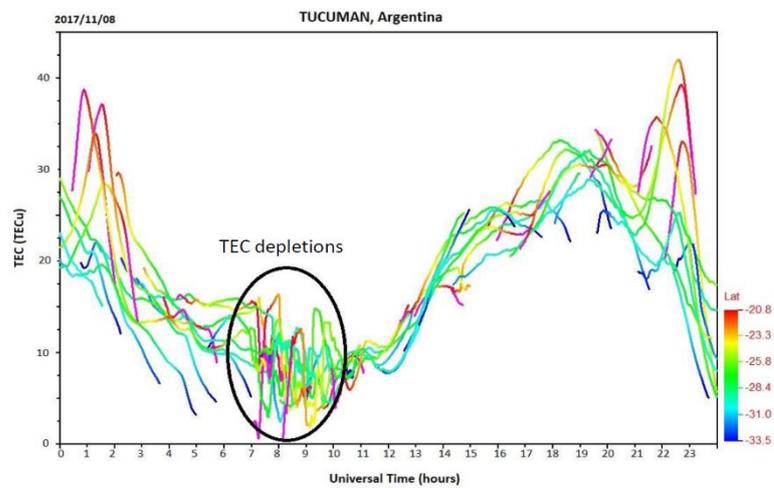

*Figure 24. TEC calibrated from a GPS receiver at Tucuman. TEC depletions (indicated with a black circle) are observed on November 8 at ~7 – 10 UT (4 – 7 LT).*

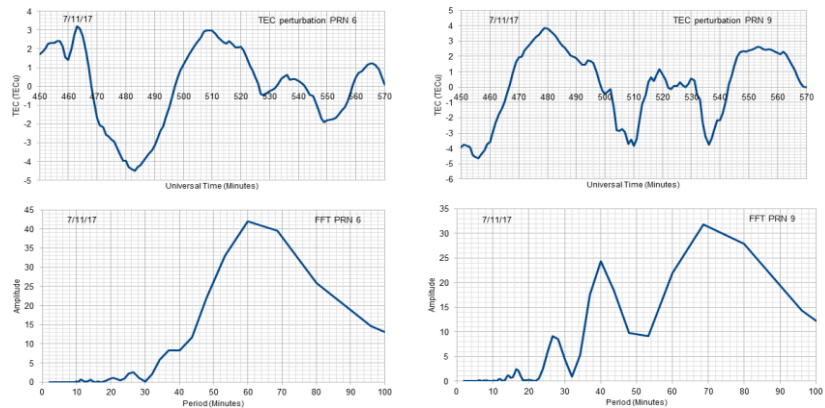

*Figure 25. Upper panel, the TEC perturbations characterizing PRNs 6 (left) and 9 (right) on November 7, 2017. Bottom panel, FFT of TEC perturbations characterizing PRNs 6 (left) and 9 (right)*

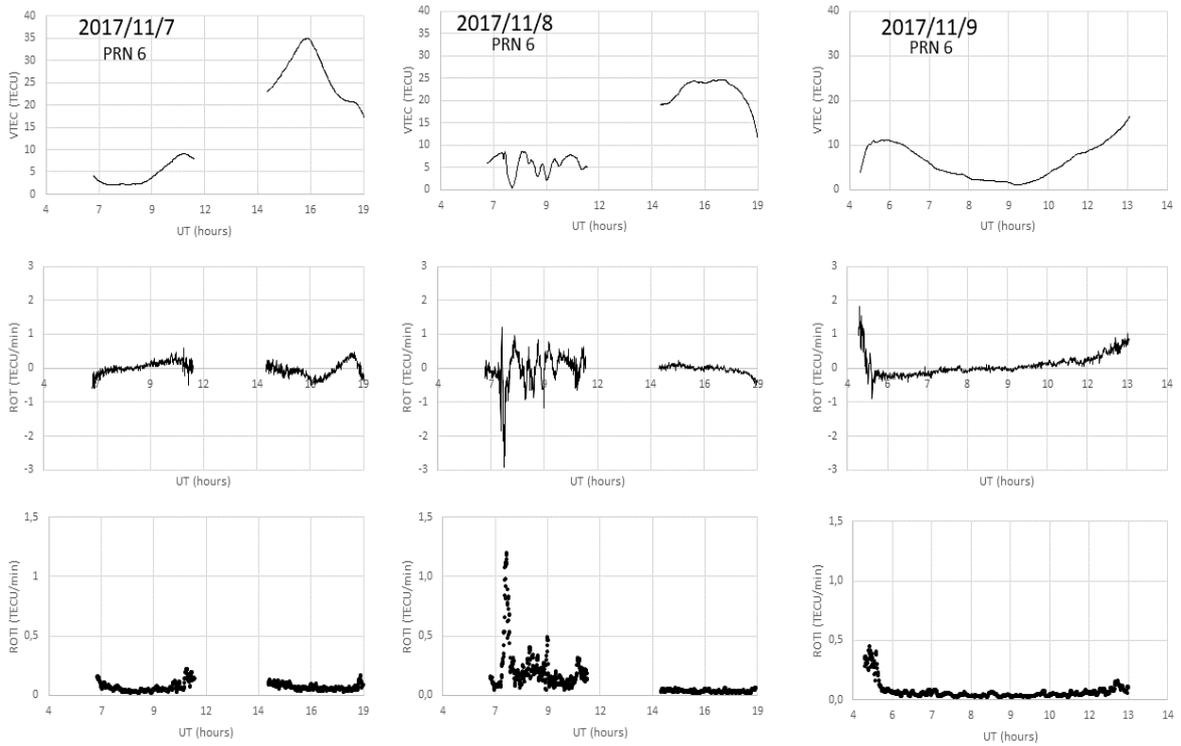

*Figure 26a. Variations of VTEC, ROT and ROTI for PRN 6 over Tucumán during November 7 – 9, 2017.*

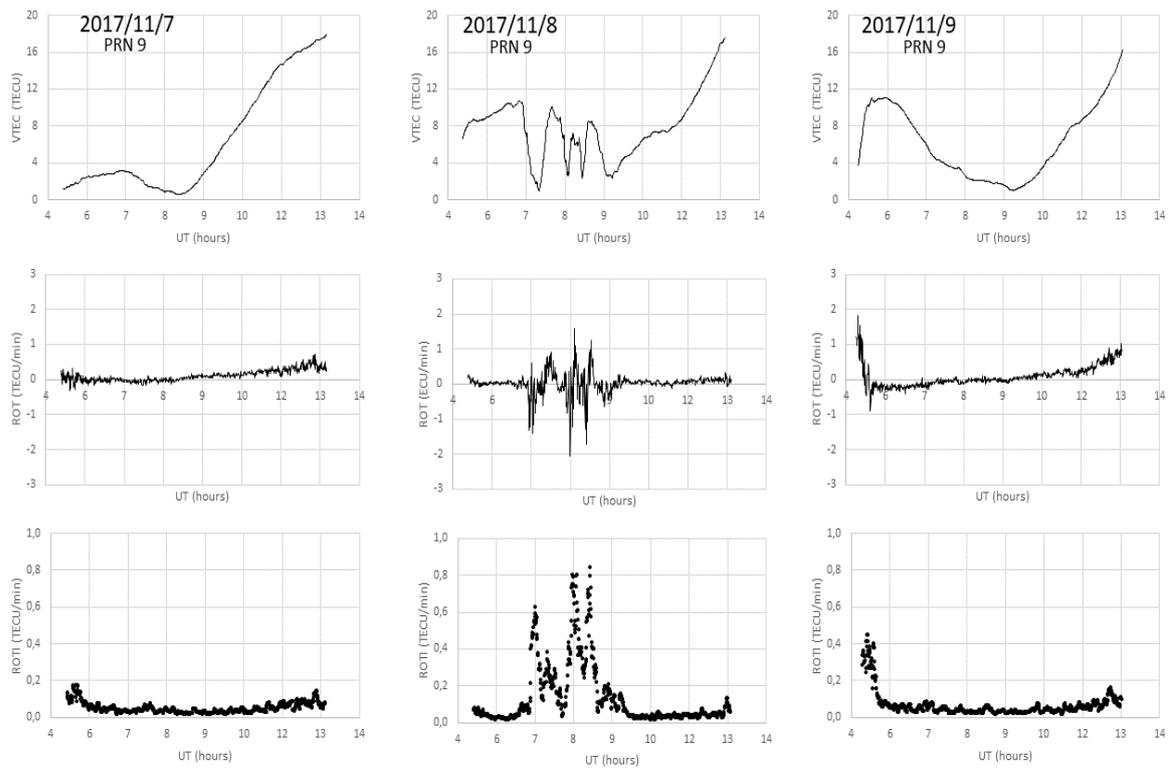

*Figure 26b. Variations of VTEC, ROT and ROTI for PRN 9 over Tucumán during November 7 – 9, 2017.*

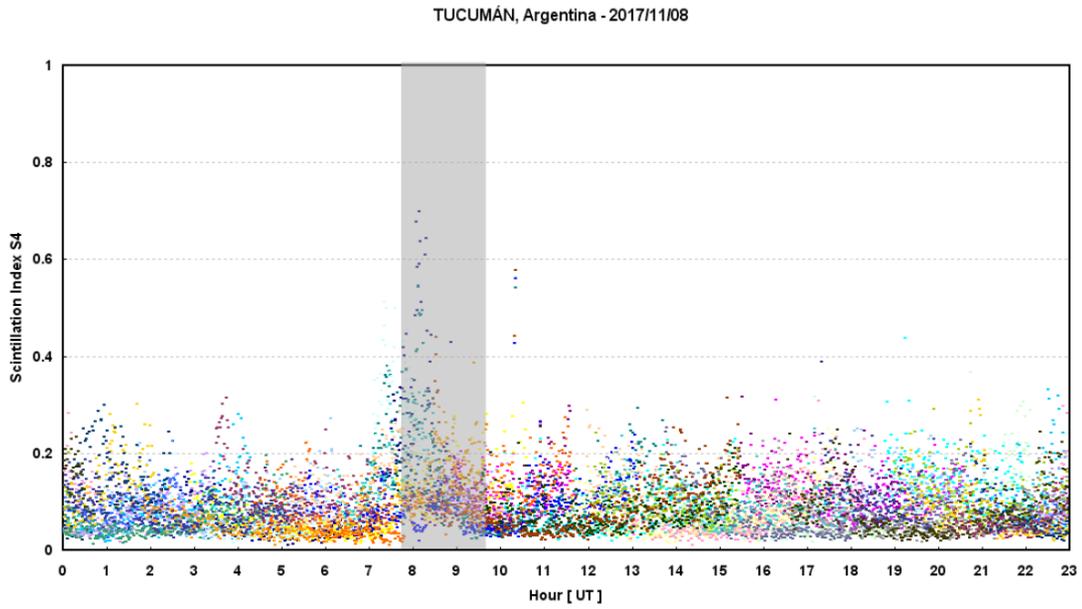

*Figure 27. The temporal variation of Global Positioning System L-band scintillations over Tucuman on November 8, 2017. The shaded region indicates the period when SSF is observed in ionograms.*

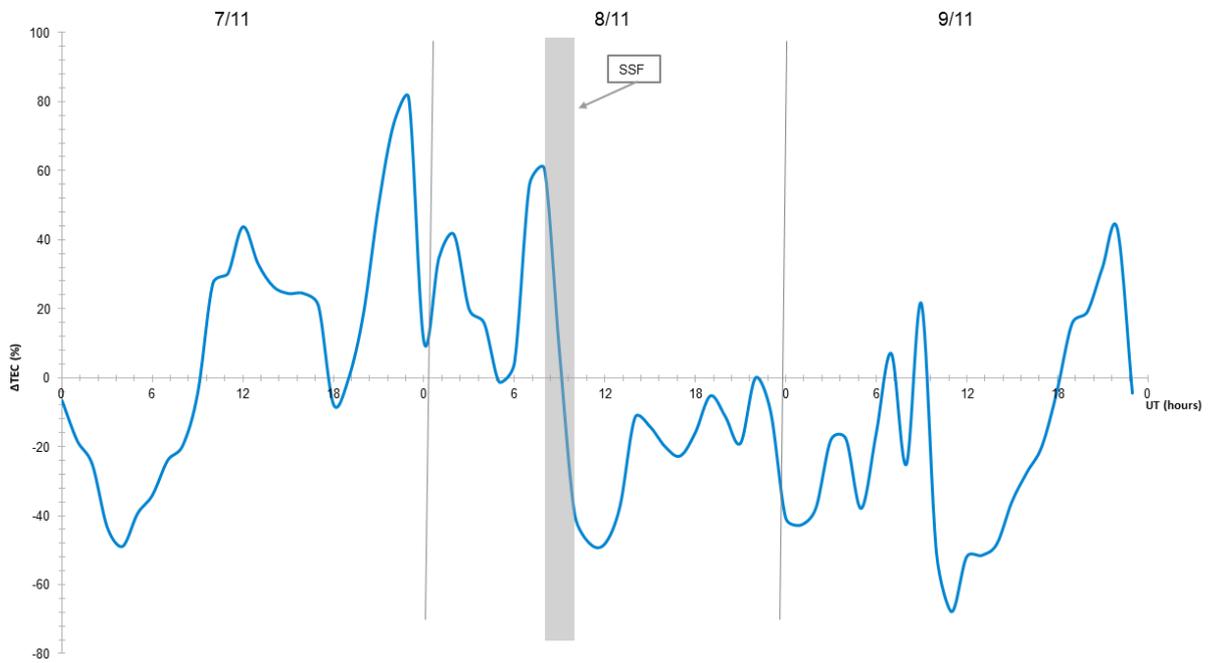

*Figure 28. Deviation ΔTEC between the TEC values for 7 – 9 November and the average TEC of the 10 quietest days of November 2017. The shaded region indicates the periods when spread-F is observed in ionograms.*

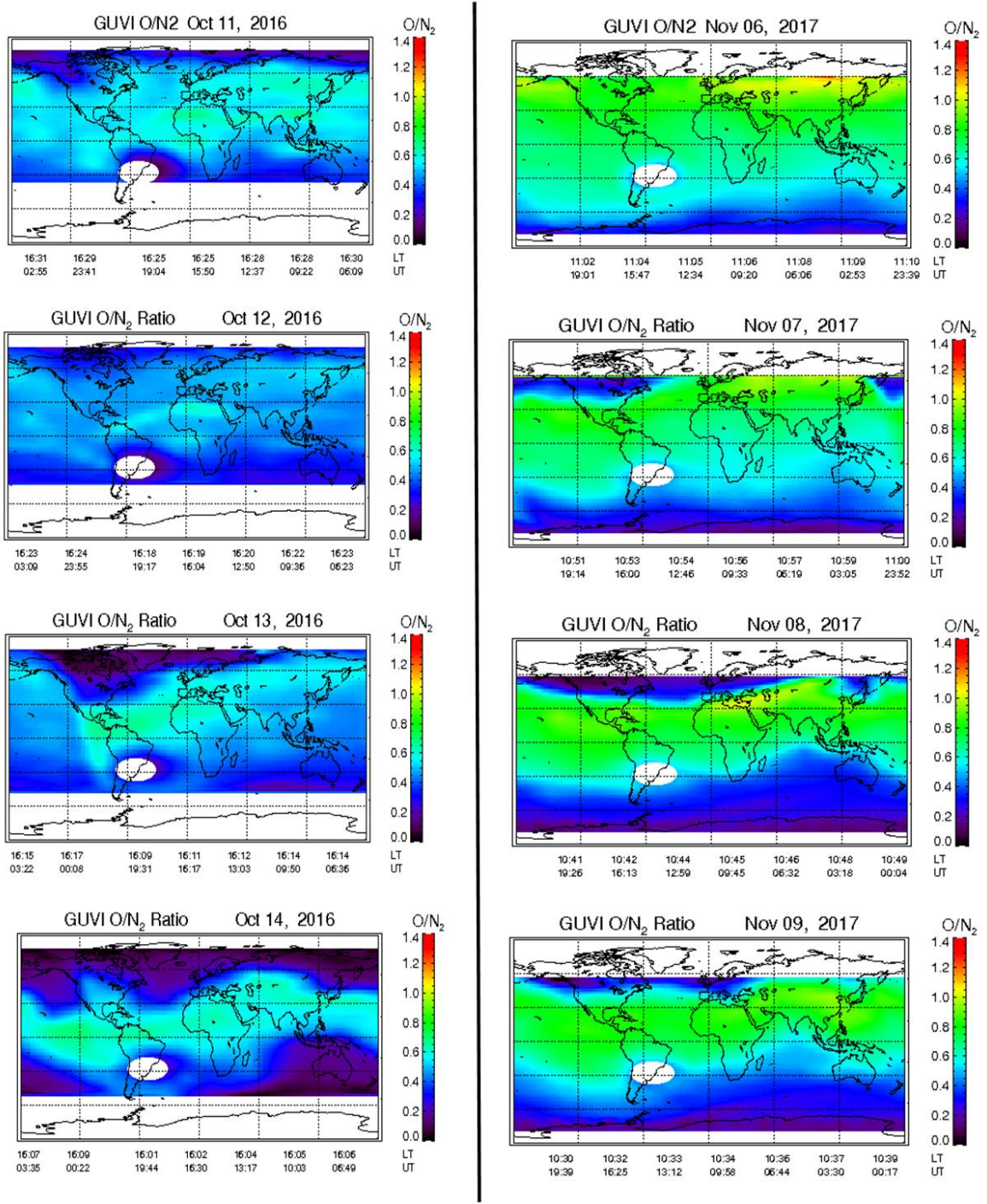

*Figure 29: Maps with the thermospheric O/N2 ratio derived from TIMED/GUVI during October 11-14, 2016 (left) and November 6-9, 2017 (right).*

As can be observed in figure 21, three periods with negative perturbations in ΔH are present on November 7 at 9-14 UT and November 8 at 9-17 UT and a weaker one on November 9 at 9-14 UT. An oscillatory behavior is present in ΔH on November 7, 9-19 UT. During the period when ionospheric irregularities are observed, ΔH was slightly negative.

Spread-F is identified from the ionograms recorded during November 2017 at Tucuman (figure 22). For this period there is no data for several days because the ionosonde was not operating. The presence of RSF was observed in six of the eighteen days available, three of them are the most disturbed days of the month. During the period of the storm, spread-F occurred on November 8 at 7:43 – 9:48 UT (4:43 – 6:48 LT), during the recovery phase of the storm. The spread-F echo extends well past the local foF2 value (i.e., foF2 is ~10 MHz whereas the trace in figure 23 extends to ~15 MHz) until 8:48 UT.

Figure 24 shows TEC vs UT on November 8, the black circle TEC depletions between 7 – 10 UT (4 – 7 LT). This coincides with the strong spread-F observed in the ionograms. The FFT analysis (figure 25) shows that the dominant periods are ~60 and 40 minutes.

VTEC, ROT and ROTI for PRN 6 and PRN 9 on November 7-9 are shown in figures 26 a and b. On November 8 at 7-9 UT, VTEC for both PRNs present TEC depletions. During this period, for PRN 6 ROT level is ~ 1 TECU/min with a peak of -3 TECU/min at 7:47UT and ROTI values are 0.4 – 1.2 TECU/min. And for PRN 9, ROT levels are ~1.5 TECU/min and ROTI values are 0.4 - 0.8 TECU/min. These characterize moderate TEC fluctuations.

Moreover, S4 index is higher than 0.5 during November 8 at ~7 – 10 UT, indicating strong scintillation activity (figure 27). In this case as in the previous one, the ionospheric irregularities that produce SSF also cause scintillation.

Figure 28 shows $\Delta TEC$ vs UT for November 7 – 9, 2017. A negative disturbance is observed during the initial phase of the storm. This is followed by irregular positive disturbances during the main phase and part of the recovery phase with a peak of ~ 84% at 22 UT (19 LT) on November 7. After ~9 UT (6 LT) on November 8, negative disturbances are present and the maximum decrease in TEC is of 78% at 11 UT (8 LT) on November 9.

## 4. Discussion

The occurrence of ionospheric irregularities at the low latitude station of Tucuman, Argentina during three geomagnetic storms are discussed using data from GPS receivers and ionosondes. To the authors' knowledge, this is the first work on ionospheric irregularities during geomagnetic storms in the region of Tucumán.

The storms occur in three different seasons; winter (low occurrence rates of PBs), equinox (transition from low to high occurrence rates of PBs) and summer (high occurrence rates of PBs). The Total electron content (TEC) estimated with a GPS-TEC calibration technique, GPS Ionospheric L-band scintillation, the virtual height of the F-layer bottom side (h'F) and the critical frequency of the F2- layer (foF2) scaled from the ionograms, are considered. Interplanetary data were used to characterize the magnetic storm phases.

For the storm that occurred in winter, RSF developed at nighttime (10 hours after the SSC) in coincidence with a positive ionospheric storm effect which was likely associated with the uplifting of the F-region. TEC depletions with periodicity of ~70 and ~40 minutes were observed, and moderate TEC fluctuations were present according to the ROTI values. At the time RSF was observed in the ionograms an eastward DDEF arising from the auroral heating and an eastward PEF associated to Bz turning north were affecting the low latitude region.

For the storm which occurred near equinox, SSF was observed at the nighttime during the initial phase of the storm (~5 hours after the SSC) and at dawn during the recovery phase, simultaneously with a positive storm effect. In the first case, no significant disturbance in AE or IMF Bz was observed. The spread-F could have been caused by upward propagating atmospheric gravity waves. Strong TEC fluctuations (ROTI ≥ 1) were observed in coincidence with ionogram spread-F, the FFT analysis of the perturbations shows periodicity of ~ 48 and 34 minutes. These TEC depletions are associated with the ionogram RSF and are a manifestation of PBs. Furthermore, unlike the previous event, strong scintillation activity occurred almost simultaneously with the ionosonde spread-F observations on October 14 and during the initial phase of the storm on October 13.

Finally, the storm occurred in summer was different to the two previous storms since it was caused by HSSWS and not by a CME. The IMF Bz polarization and $E_y$ oscillated rapidly simultaneously with a decreased in Dst and an irregular increased in the AE index. At dawn during the recovery phase, SSF was present in the ionograms in addition to strong scintillation activity and moderate TEC fluctuation with periodicity of ~60 and ~40 minutes.

The large increase observed in AE index during the storms of May 27 and October 12 is an indication of energy and momentum deposition into the high-latitude ionosphere that produce auroral heating. As several researchers have reported (Blanc & Richmond, 1980; Scherliess & Fejer, 1997; Senior & Blanc, 1984), this generates thermospheric disturbance winds that can drive DDEF affecting low latitudes several hours after the SSC. This eastward electric field produces an upward disturbance vertical drift in the F region. This is indicated with a rapid F layer height rise that generates an unstable plasma density profile. Further, this leads to the development of spread-F irregularities through the RTI process, even during a season of minimal spread-F occurrence, like May.

As it was mention before, the disturbance winds and the associated dynamo electric field take a few hours to set up. Therefore, it is reasonable to associate the ionospheric effects observed several hours after the storm main phase onset with DDEF. Disturbance dynamo processes were likely acting during the storm recovery phase on November 8, and may be added to the PPEFs associated to the oscillatory behavior in IMF Bz. A possibility is that during the night the eastward disturbance dynamo electric field elevated the F layer and favored the generation of spread-F. Furthermore, the analysis of ΔH suggests the presence of a westward prompt penetration electric field which was likely associated with Bz turning north that disrupted the development of these irregularities.

The uplifting of the F layer to heights where fewer molecular species are present could be responsible for the large positive ionospheric storms observed during the three events studied. Furthermore, changes in the neutral composition of the upper atmosphere could also play an important role in the ionospheric ion density distribution during geomagnetic storms (Fuller-Rowell, Codrescu, Moffett, & Quegan, 1994). A possible correlation between increases in the thermospheric O/N2 ratio and positive ionospheric storms have been reported (Astafyeva et al., 2018; Mansilla & Zossi, 2020). Figure 29 shows the variations in the O/N2 ratio during the geomagnetic storms occurred in October 2016 and November 2017. These data were obtained from the Global Ultraviolet Imager (GUVI) on board NASA's Thermosphere, Ionosphere, and Mesosphere, Energetics and Dynamics (TIMED) satellite (http://guvitimed.jhuapl.edu/). Tucumán is close to the South Atlantic Magnetic Anomaly (SAMA) and there are no data under this region. Nevertheless, the behavior around the SAMA region can be used to infer the variations in the O/N2 ratio over Tucumán. For May 28 and 29, there are no data for latitudes south of ~15°S so the analysis is limited to the storms of October and November. Enhancements in the O/N2 ratio are observed in coincidence with positive storm effect on October 13, 2016. In contrast, on November 7, 2017 when a positive effect is observed during the main phase of the storm, O/N2 ratio slightly decreases around the SAMA region. Thus, in the present work, no correlation was found between the increase in the thermospheric O/N2 ratio and positive ionospheric storm effects.

Several authors (M. a. Abdu, 2012; Aquino & Sreeja, 2013; Basu et al., 2001; Bhattacharyya, Basu, Groves, Valladares, & Sheehan, 2002; Huang, 2011; Stanislawska, Lastovicka, Bourdillon, Zolesi, & Cander, 2010) have shown that a geomagnetic storm could act as an inhibitor or as an initiator of ionospheric irregularities, depending on changes in the quiet and disturbed drift patterns during different seasons. Becker-Guedes et al. (2004) discussed three case studies at Brazilian stations and found that during low PBs occurrence season and transition season geomagnetic activity contributes to the generation of irregularities, while inhibiting them in the high PBs occurrence season. Sahai et al. (2007) reported that for two stations in the Brazilian sector during an intense geomagnetic storm in August 2003, spread-F was observed during the recovery phase in the nighttime. On the other hand, de Abreu et al. (2017) studied the effects of an intense geomagnetic storm over the American sector, they observed that the storm did not influence the generation or suppression of ionospheric irregularities.

The present work shows that, for the three storms occurred in different seasons, geomagnetic activity creates favorable conditions for the initiation of ionospheric irregularities, manifested as ionogram spread-F and TEC wave-like fluctuation. The occurrence of PBs during the geomagnetic storms analyzed here is related to the upward movement of F region resulting from eastward electric field perturbations. These observations are in agreement with Tulasi Ram et al. (2008), who pointed out that the local time dependence of the polarity and amplitude of electric field perturbations (PPEF and DDEF) during geomagnetically active periods determines the favorable or unfavorable conditions for the development of spread-F irregularities through the growth of the RTI process. Additionally, Abdu et al. (2012) showed that for three stations Sao Luis, Fortaleza, and Jicamarca, the F layer rise due to the DDEF was followed with spread F developed at nighttime (21-3 LT) during an intense storm period. Recently, de Paula et al. (2019) studied the ionospheric irregularity over São Luís, Brazil during the two-step magnetic storm of September 6 –10, 2017. They found that an under-shielding eastward electric field caused a large upward plasma drift during the time of the evening pre-reversal vertical drift on September 7, which triggered strong scintillation during the post-sunset hours. While westward DDEF was suggested to be the cause of a downward movement of the F layer height and the scintillation inhibition on September 8. Sahai et al. (2011) reported the inhibition of the formation of post-sunset spread-F in the Latin American sector during the intense geomagnetic storm of January 21 2005 due to DDEF. Cherniak et al. (2019) show the presence of post-sunset PBs in the equatorial ionosphere induce by PPEF during the intense geomagnetic storm of June 22-23, 2015 for the period of lowest PBs occurrence.

The ROT and ROTI index are used in this work to describe the intensity of ionospheric TEC fluctuations. For the storms presented here it is observed that ROT fluctuations and high ROTI values coincides with TEC depletions. ROTI ≥ 0.5 TECU/min corresponds with periods when fluctuations are observed in TEC, this indicate the presence of ionospheric irregularities of several kilometers (Ma and Maruyama, 2006). In the present analysis Five minute window is used to calculate ROTI, as it is explained by Nishioka et al. (2008) this method detects irregularities of ~ 20 km of spatial scale. Therefore, ROTI identifies the substructures inside the plasma bubbles.

Ngwira et al. (2013) studied the ionospheric response during a minor geomagnetic storm and Amaechi et al. (2018) investigated the effects of four intense geomagnetic storms on the occurrences of ionospheric irregularities over the African low-latitude region. Both works used TEC measurements (TEC perturbation, ROT and ROTI) to examine the presence of ionospheric irregularities. They found that high values of ROTI correspond to periods of electron density depletions/fluctuations associated with equatorial plasma bubbles. The same behavior is observed in Tucumán in the present work. Liu et al. (2016) utilized ROTI to analyzed the characteristics of TEC fluctuation over China. They considered ROTI ≥ 0.5 indicates the occurrence of irregular ionospheric activities relevant to ionospheric scintillation. Our results show strong amplitude scintillation activity in coincidence with moderate TEC fluctuation except for May 28.

Valladares et al. (2004) defined a TEC depletion as a sudden reduction of TEC followed by a recovery to a level near the TEC value preceding the depletion. As it has been explained by several researchers (DasGupta A. et al., 1983; Dashora & Pandey, 2005; Tsunoda & Towle, 1979; Weber et al., 1996), the TEC depletions are a manifestation of plasma bubbles that drift across the line of-sight between the GPS receiver and the satellite.

Plasma bubbles rise to great heights in the magnetic equator and drift along the magnetic field line to the anomaly crest. At the edges of the plasma bubble, the steep density gradients could create favorable conditions to the generation of small-scale irregularities (hundreds of meters) that induce GPS scintillation (Muella et al., 2010; Ray, Paul, & Dasgupta, 2006). Previous works had reported a good correlation between TEC depletions and strong scintillation (Bagiya & Sridharan, 2011; Dashora & Pandey, 2005; Olwendo, Cilliers, Baki, & Mito, 2012; Seemala & Valladares, 2011). The data presented here shows correspondence between TEC depletions and amplitude scintillation for the storms of October and November but not for May when S4 was generally less than 0.2.

Some researchers used a depth threshold of 5 TECU to consider a TEC depletion to be related to bubbles (Magdaleno, Herraiz, & de la Morena, 2012; Shetti, Gurav, & Seemla, 2019). In the present work, TEC depletions with depth of 3 – 15 TECU were observed. The shallowest depletions occurred in May 28 coinciding with weak scintillation activity and moderate ROTI. Deng et al.,(2015) observed that TEC depletions with depth smaller than 10 TECU were associated with small or moderate ROTI and with weak or no scintillation in the region of the northern crest of the EIA over China. They concluded that eroded plasma bubbles containing large-scale ROTI irregularities and the disappearance or decay of small-scale irregularities may be responsible for these depletions. This could explain the observations for the storm of May.

## 5. Conclusions

This work presents the first report on the generation/suppression of ionospheric irregularities in the region of Tucumán, Argentina during geomagnetic storms. The storms studied occurred on May 27, 2017 (a month of low occurrence rates of spread – F), October 12, 2016 (a month of transition from low to high occurrence rates of spread-F) and November 7, 2017 (a month of high occurrence rates of spread-F).

We suggest that in all cases eastward DDEF may be responsible for the generation of post-midnight irregularities. These irregularities are manifested in the form of fluctuations in TEC and spread F in ionograms. For the storm in May, an eastward over-shielding PPEF seems to be present during the final part of the main phase and the beginning of the recovery phase. A possibility is that the PPEF is added to the DDEF and produces a rise in the F region that is favorable to irregularity generation. Irregularities were generally observed during the main and recovery phases of the storm. Moreover, during the storms of October and November, strong GPS L band scintillation is observed associated with strong spread-F (SSF).

More studies are needed towards improving the understanding of the coupling processes that control the irregularity occurrence/suppression under disturbed geomagnetic conditions, such as the magnetosphere–ionosphere coupling that cause perturbation electric fields and winds. These are key parameters to describe and model the short-term variability of the ionospheric weather.

**Acknowledgments** Special thanks to Prof. Otsuka from Nagoya University, Dr. Valladares from UT. Dallas and Prof. Mansilla from National University of Tucumán for their helpful suggestions. The author is also thankful to the following groups for making data publicly available: the Low Latitude Ionospheric Sensor Network (LISN), the upper atmosphere physics group of the Istituto Nazionale di Geofisica e Vulcanologia – Italy, the National Geographic Institute (IGN), the UMass Lowell Center for Atmospheric Research, the NASA/Goddard's Space Physics Data Facility, the International Service of Geomagnetic Indices, the GFZ German Research Center for Geosciences at Potsdam and the World Data Center (WDC) Kyoto, Japan.

**Funding.** This work was supported by Universidad del Norte Santo Tomás de Aquino (UNSTA).

**References**

Abdu, M. a. (2012). Equatorial spread F/plasma bubble irregularities under storm time disturbance electric fields. *Journal of Atmospheric and Solar-Terrestrial Physics*, *75–76*, 44–56. http://doi.org/10.1016/j.jastp.2011.04.024

Abdu, M. A., Batista, I. S., Bertoni, F., Reinisch, B. W., Kherani, E. A., & Sobral, J. H. A. (2012). Equatorial ionosphere responses to two magnetic storms of moderate intensity from conjugate point observations in Brazil. *Journal of Geophysical Research: Space Physics*, *117*(5), 1–20. http://doi.org/10.1029/2011JA017174

Abdu, M. A., de Medeiros, R. T., Sobral, J. H. A., & Bittencourt, J. A. (1982). Spread F plasma bubble vertical rise velocities determined from spaced ionosonde observations. *Secretaria de Planejamiento Da Presidencia Da Republica*, (INPE-2416-PRE/124), 25.

Alfonsi, L. *et al.* (2013) 'Comparative analysis of spread-F signature and GPS scintillation occurrences at Tucumán, Argentina', *Journal of Geophysical*

Research: Space Physics, 118(7), pp. 4483–4502. doi: 10.1002/jgra.50378.

Amaechi, P. O., Oyeyemi, E. O., & Akala, A. O. (2018). Geomagnetic storm effects on the occurrences of ionospheric irregularities over the African equatorial/low-latitude region. *Advances in Space Research*, *61*(8), 2074–2090.
https://doi.org/10.1016/j.asr.2018.01.035

Aquino, M., & Sreeja, V. (2013). Correlation of scintillation occurrence with interplanetary magnetic field reversals and impact on global navigation satellite system receiver tracking performance. *Space Weather*, *11*(5), 219–224. http://doi.org/10.1002/swe.20047

Astafyeva, E., Zakharenkova, I., Hozumi, K., Alken, P., Coïsson, P., Hairston, M. R., & Coley, W. R. (2018). Study of the Equatorial and Low-Latitude Electrodynamic and Ionospheric Disturbances During the 22-23 June 2015 Geomagnetic Storm Using Ground-Based and Spaceborne Techniques. *Journal of Geophysical Research: Space Physics*, *123*(3), 2424–2440. http://doi.org/10.1002/2017JA024981

Atıcı, R., & Sağır, S. (2019). Global investigation of the ionospheric irregularities during the severe geomagnetic storm on September 7–8, 2017. *Geodesy and Geodynamics*, (June), 1–11.
https://doi.org/10.1016/j.geog.2019.05.004

Azzouzi, I., Migoya-Orue, Y. O., Coisson, P., Amory-Mazaudier, C., Fleury, R., & Radicella, S. M. (2016). Day-to-day variability of VTEC and ROTI in October 2012 with impact of high-speed solar wind stream on 13 October 2012. *Sun and Geosphere*, *11*(October 2012), 7–22.

Basu, S., K. M. Groves, J. M. Quinn, and P. Doherty (1999), A comparison of TEC fluctuations and scintillations at Ascension Island, J. Atmos. Sol. Terr. Phys., 61, 1219–1226

Basu, S., Basu, S., Valladares, C. E., Yeh, H.-C., Su, S.-Y., MacKenzie, E., … Bullett, T. W. (2001). Ionospheric effects of major magnetic storms during the International Space Weather Period of September and October 1999: GPS observations, VHF/UHF scintillations, and in situ density structures at middle and equatorial latitudes. *Journal of Geophysical Research*, *106*(A12), 30389. http://doi.org/10.1029/2001JA001116

Batista, I. S., & Abdu, M. A. (1977). Magnetic storm associated delayed sporadic E enhancements in the Brazilian Geomagnetic Anomaly. *Journal of Geophysical Research*, *82*(29), 4777–4783. http://doi.org/10.1029/JA082i029p04777

Becker-Guedes, F., Sahai, Y., Fagundes, P. R., Lima, W. L. C., Pillat, V. G., Abalde, J. R., & Bittencourt, J. A. (2004). Geomagnetic storm and equatorial spread-F. *Annales Geophysicae*, (22), 3231–3239. http://www.ann-geophys.net/22/3231/2004/angeo-22-3231-2004.pdf

Bhattacharyya, A., Basu, S., Groves, K. M., Valladares, C. E., & Sheehan, R. (2002). Effect of magnetic activity on the dynamics of equatorial F region irregularities. *Journal of Geophysical Research: Space Physics*, *107*(A12), SIA 20-1-SIA 20-7. http://doi.org/10.1029/2002JA009644

Blanc, M. and Richmond, A. D. (1980) 'The ionospheric disturbance dynamo', *Journal of Geophysical Research*, 85(A4), pp. 1669–1686. doi: 10.1029/JA085iA04p01669.

Bowman, G. G. (1991). Ionospheric Features Associated with the Fine Structure Recorded on Some Mid-Latitude Spread-F Ionograms. *J. Geomag. Geoelectr.*, *43*, 885–897.

Bullett, T. (2008). *Station Report : A new ionosonde at Boulder*. Cooperative Institute

for Research in Environmental Sciences. University of Colorado, Boulder.

Calvert, W. (1962). *Equatorial Spread-F*. Technical note. US department of Commerce. National Bureau of Standards.

Cherniak, I., Zakharenkova, I. and Sokolovsky, S. (2019) 'Multi-instrumental observation of storm-induced ionospheric plasma bubbles at equatorial and middle latitudes', *Journal of Geophysical Research: Space Physics*, (2), pp. 0–2. doi: 10.1029/2018JA026309.

Davies. K., (1990). Ionospheric Radio, Peter Peregrinus Ltd., London, U.K., 291 pp.

de Abreu, A. J., Martin, I. M., Fagundes, P. R., Venkatesh, K., Batista, I. S., de Jesus, R., … Wild, M. (2017). Ionospheric F-region observations over American sector during an intense space weather event using multi-instruments. *Journal of Atmospheric and Solar-Terrestrial Physics*, *156*, 1–14. https://doi.org/10.1016/j.jastp.2017.02.009

de Medeiros, R. T., Abdu, M. A. and Kantor, I. J. (1983) 'A comparative study of VHF scintillation and spread F events over Natal and Fortaleza in Brazil', *Journal of Geophysical Research*, 88(A8), p. 6253. doi: 10.1029/JA088iA08p06253.

de Paula, E. R., de Oliveira, C. B. A., Caton, R. G., Negreti, P. M., Batista, I. S., Martinon, A. R. F., … Moraes, A. O. (2019). Ionospheric irregularity behavior during the September 6–10, 2017 magnetic storm over Brazilian equatorial–low latitudes. *Earth, Planets and Space*, *71*(1), 42. https://doi.org/10.1186/s40623-019-1020-z

Dugassa, T., Habarulema, J. B., & Nigussie, M. (2020). Statistical study of geomagnetic storm effects on the occurrence of ionospheric irregularities over equatorial/low-latitude region of Africa from 2001 to 2017. *Journal of Atmospheric and Solar-Terrestrial Physics*, 105198. https://doi.org/10.1016/j.jastp.2020.105198

Fejer, B. G., Blanc, M., & Richmond, A. D. (2017). Post-Storm Middle and Low-Latitude Ionospheric Electric Fields Effects. *Space Science Reviews*, *206*(1–4), 407–429. http://doi.org/10.1007/s11214-016-0320-x

Fejer, B. G., Jensen, J. W., & Su, S.-Y. (2008). Seasonal and longitudinal dependence of equatorial disturbance vertical plasma drifts. *Geophysical Research Letters*, *35*(20), L20106. http://doi.org/10.1029/2008GL035584

Fejer, B. G., Larsen, M. F., & Farley, D. T. (1983). Equatorial disturbance dynamo electric fields. *Geophysical Research Letters*, *10*(7), 537–540. http://doi.org/10.1029/GL010i007p00537

Gonzalez, W. D., Tsurutani, B. T., & Clúa De Gonzalez, A. L. (1999). Interplanetary origin of geomagnetic storms. *Space Science Reviews*, *88*(3–4), 529–562. https://doi.org/10.1023/A:1005160129098

Huang, C. S. (2011). Occurrence of equatorial plasma bubbles during intense magnetic storms. *International Journal of Geophysics*, *2011*, 1–10. http://doi.org/10.1155/2011/401858

Hines, C. O. (1959), An interpretation of certain ionospheric motions in terms of atmospheric gravity waves, J. Geophys. Res., 64, 2210–2211, doi:10.1029/JZ064i012p02210.

Hooke WH (1968) Ionospheric irregularities produced by internal atmospheric gravity waves. J Atmos Terr Phys 30:795–823

Hunsucker RD (1982) Atmospheric gravity waves generated in the high latitude ionosphere: a review. Rev Geophys 20:293–315. https://doi.org/10.1029/RG020i002p00293

Jayachandran, P. T., Ram, P. S., Somayajulu, V. V, & Rao, P. V. S. R. (1997). Effect of equatorial ionization anomaly on the occurrence of spread-F, *262*, 255–262.

Kirchengast G, Hocke K, Schlegel K (1996) The gravity wave-TID relationship: insight via theoretical model-EISCAT data comparison. J. Atmos. Sol. Terr. Phys. 58:233–243. https://doi.org/10.1016/0021-9169(95)00032-1

Liu, X., Yuan, Y., Tan, B., & Li, M. (2016). Observational Analysis of Variation Characteristics of GPS-Based TEC Fluctuation over China. *ISPRS International Journal of Geo-Information*, *5*(12), 237. https://doi.org/10.3390/ijgi5120237

Lyon, A. J., Skinner, N. J., & Wright, R. W. H. (1960). The belt of equatorial spread-F. *Journal of Atmospheric and Terrestrial Physics*, *19*(3–4), 145–159. http://doi.org/10.1016/0021-9169(60)90043-X

Ma, G., & Maruyama, T. (2006). A super bubble detected by dense GPS network at east Asian longitudes. *Geophysical Research Letters*, *33*(21), L21103 http://doi.wiley.com/10.1029/2006GL027512

Mansilla, G. A. (2001). Thermosphere-ionosphere disturbances during two intense geomagnetic storms. *Geofísica Internacional*, *40*(2), 135–144.

Martinis, C. R. (2005). Toward a synthesis of equatorial spread F onset and suppression during geomagnetic storms. *Journal of Geophysical Research*, *110*(A7), A07306. http://doi.org/10.1029/2003JA010362

Ngwira, C. M., Seemala, G. K., & Bosco Habarulema, J. (2013). Simultaneous observations of ionospheric irregularities in the African low-latitude region. *Journal of Atmospheric and Solar-Terrestrial Physics*, *97*, 50–57. https://doi.org/10.1016/j.jastp.2013.02.014

Nishioka, M., Saito, A., & Tsugawa, T. (2008). Occurrence characteristics of plasma bubble derived from global ground-based GPS receiver networks. *Journal of Geophysical Research: Space Physics*, *113*(A5), n/a-n/a. https://doi.org/10.1029/2007JA012605

Otsuka, Y., Shiokawa, K. and Ogawa, T. (2006) 'Equatorial Ionospheric Scintillations and Zonal Irregularity Drifts Observed with Closely-Spaced GPS Receivers in Indonesia', *Journal of the Meteorological Society of Japan*, 84A, pp. 343–351. doi: 10.2151/jmsj.84A.343.

Pavlov, A. V., Fukao, S., Kawamura, S., Propagation, R., Region, M., & Technology, C. (2006). A modeling study of ionospheric F2-region storm effects at low geomagnetic latitudes during 17 – 22 March 1990. *Annales Geophysicae*, *24*(3), 915–940. http://doi.org/10.5194/angeo-24-915-2006

Pezzopane M, Scotto C (2007) The automatic scaling of critical frequency foF2 and MUF(3000)F2: a comparison between Autoscala and ARTIST 4.5 on Rome data. Radio Sci. https://doi.org/10.1029/2006RS003581

Pi, X., Mannucci, A. J., Lindqwister, U. J., & Ho, C. M. (1997). Monitoring of global ionospheric irregularities using the Worldwide GPS Network. *Geophysical Research Letters*, *24*(18), 2283–2286. https://doi.org/10.1029/97GL02273

Piggott, W. R., and Rawer K. (1972), URSI handbook of ionogram interpretation and reduction, Rep. UAG-23A, pp. 68–73, World Data Cent. A for Sol.-Terr. Phys., Boulder, Colo

Piñón, D. A., Gómez, D. D., Smalley, R., Cimbaro, S. R., Lauría, E. A. and Bevis, M. G. (2018) 'The History, State, and Future of the Argentine Continuous Satellite Monitoring

Network and Its Contributions to Geodesy in Latin America', *Seismological Research Letters*, 89(2A), pp. 475–482. doi: 10.1785/0220170162.

Rastogi, R. G. *et al.* (1989) 'Spread-F and radio wave scintillations near the F-region anomaly crest', *Annales Geophysicae*, 7, pp. 281–284.

Ray, S., Roy, B., & Das, A. (2015). Occurrence of equatorial spread F during intense geomagnetic storms. *Radio Science*, *50*(7), 563–573. http://doi.org/10.1002/2014RS005422

Sahai, Y., Becker-Guedes, F., Fagundes, P. R., Lima, W. L. C., Otsuka, Y., Huang, C.-S., … Bittencourt, J. A. (2007). Response of nighttime equatorial and low latitude F-region to the geomagnetic storm of August 18, 2003, in the Brazilian sector. *Advances in Space Research*, *39*(8), 1325–1334. https://doi.org/10.1016/j.asr.2007.02.064

Sahai, Y., Fagundes, P. R., de Jesus, R., de Abreu, A. J., Crowley, G., Kikuchi, T., … Bittencourt, J. A. (2011). Studies of ionospheric F-region response in the Latin American sector during the geomagnetic storm of 21-22 January 2005. *Annales Geophysicae*, *29*(5), 919–929. https://doi.org/10.5194/angeo-29-919-2011

Scherliess, L. and Fejer, B. G. (1997) 'Storm time dependence of equatorial disturbance dynamo zonal electric fields', *Jounal of Geophysical Research*, 102(97), pp. 24037–24046.

Senior, C. and Blanc, M. (1984) 'On the Control of Magnetospheric Convection by the Spatial Distribution of Ionospheric Conductivities', *Journal of Geophysical Research*, 89(A1), pp. 261–284. doi: 10.1029/JA089iA01p00261.

Shi, J. K. *et al.* (2011) 'Relationship between strong range spread F and ionospheric scintillations observed in Hainan from 2003 to 2007', *Journal of Geophysical Research: Space Physics*, 116, pp. 1–5. doi: 10.1029/2011JA016806.

Sobral, J. H. A., Abdu, M. A., Gonzalez, W. D., Tsurutani, B. T., Batista, I. S., & DeGonzalez, A. L. C. (1997). Effects of intense storms and substorms on the equatorial ionosphere/thermosphere system in the American sector from ground-based and satellite data. *Journal of Geophysical Research*, *102*(A7), 14305–14313. http://doi.org/10.1029/97ja00576

Stanislawska, I., Lastovicka, J., Bourdillon, A., Zolesi, B., & Cander, L. R. (2010). Monitoring and modeling of ionospheric characteristics in the framework of European COST 296 Action MIERS. *Space Weather*, *8*(2), 1–7. http://doi.org/10.1029/2009SW000493

Tsurutani, B. T. and Gonzalez, W. D. (1987) 'The cause of high-intensity long-duration continuous AE activity (HILDCAAs): Interplanetary Alfvén wave trains', *Planetary and Space Science*, 35(4), pp. 405–412. doi: 10.1016/0032-0633(87)90097-3.

Tulasi Ram, S. *et al.* (2008) 'Local time dependent response of postsunset ESF during geomagnetic storms', *Journal of Geophysical Research: Space Physics*, 113(A7), , A07310 doi: 10.1029/2007JA012922.

Valladares, C. E., Villalobos, J., Hei, M. A., Sheehan, R., Basu, S., Mackenzie, E., … Rios, V. H. (2009). Simultaneous observation of traveling ionospheric disturbances in the Northern and Southern Hemispheres. *Annales Geophysicae*, (27), 1501–1508.

Wang, Z., J. K. Shi, K. Torkar, G. J. Wang, and X. Wang (2014), Correlation between ionospheric strong range spread F and scintillations observed in Vanimo station, J. Geophys. Res. Space Physics, 119, 8578–8585, doi:10.1002/ 2014JA020447.

Wei, Y., Zhao, B., Li, G., & Wan, W. (2015). Electric field penetration into Earth's ionosphere: a brief review for 2000–2013. *Science Bulletin*, *60*(8), 748–761. https://doi.org/10.1007/s11434-015-0749-4